\UseRawInputEncoding
\documentclass[%
 reprint,
superscriptaddress,
showpacs,preprintnumbers,
 amsmath,amssymb,
 aps,
]{revtex4-1}

\usepackage{graphicx}
\usepackage{dcolumn}
\usepackage{bm}
\usepackage{multirow}

\usepackage{amssymb}
\usepackage{amsmath}
\usepackage{colordvi}
\usepackage[colorlinks]{hyperref}
\usepackage{amsthm}
\usepackage{subeqnarray}
\usepackage{cases}
\usepackage{enumerate}
\usepackage{graphicx}
\usepackage{epsfig}
\usepackage{epstopdf}
\usepackage{subfigure}
\usepackage{color}
\usepackage{bm}

\def\avg(#1){\langle#1\rangle}

\def\be{\begin{equation}}
\def\ee{\end{equation}}
\def\bea{\begin{eqnarray}}
\def\eea{\end{eqnarray}}

\def\nn{\nonumber}

\usepackage{ulem}

\newcommand{\blue}[1]{{\color{blue}#1}}

\begin{document}

\title{Bose-Einstein condensates in an atom-optomechanical system with effective  global non-uniform interaction}

\author{Jia-Ming Cheng}
\affiliation{CAS Key Lab of Quantum Information, University of Science and Technology of China, Hefei, 230026, P.R. China}
\affiliation{Synergetic Innovation Center of Quantum Information and Quantum Physics, University of Science and Technology of China, Hefei, 230026, P.R. China}
\affiliation{Xi'an Microelectronics Technology Institute, Xi'an, 710065, P.R. China}

\author{Zheng-Wei Zhou}
\email{zwzhou@ustc.edu.cn}
\affiliation{CAS Key Lab of Quantum Information, University of Science and Technology of China, Hefei, 230026, P.R. China}
\affiliation{Synergetic Innovation Center of Quantum Information and Quantum Physics, University of Science and Technology of China, Hefei, 230026, P.R. China}

\author{Guang-Can Guo}
\affiliation{CAS Key Lab of Quantum Information, University of Science and Technology of China, Hefei, 230026, P.R. China}
\affiliation{Synergetic Innovation Center of Quantum Information and Quantum Physics, University of Science and Technology of China, Hefei, 230026, P.R. China}

\author{Han Pu}
\email{hpu@rice.edu}
\affiliation{Department of Physics and Astronomy, and Rice Center for Quantum Materials, Rice University, Houston, TX 77251, USA}

\author{Xiang-Fa Zhou}
\email{xfzhou@ustc.edu.cn}
\affiliation{CAS Key Lab of Quantum Information, University of Science and Technology of China, Hefei, 230026, P.R. China}
\affiliation{Synergetic Innovation Center of Quantum Information and Quantum Physics, University of Science and Technology of China, Hefei, 230026, P.R. China}

\date{\today}

\begin{abstract}
We consider a hybrid atom-optomechanical system consisting of a mechanical membrane inside an optical cavity and an atomic Bose-Einstein condensate outside the cavity. The condensate is confined in an optical lattice potential formed by a traveling laser beam reflected off one cavity mirror. We derive the cavity-mediated effective atom-atom interaction potential, and find that  it is non-uniform, site-dependent, and does not decay as the interatomic distance increases.
We show that the presence of this effective interaction breaks the Z$_2$ symmetry of the system and gives rise to new quantum phases and phase transitions.
When the long-range interaction dominates, the condensate breaks the translation symmetry and turns into a novel self-organized lattice-like state with increasing particle densities for sites farther away from the cavity.	
We present the phase diagram of the system, and investigate the stabilities of different phases by calculating their respective excitation spectra.
The system can serve as a platform to explore various self-organized phenomena induced by the long-range interactions.
\end{abstract}

\maketitle

\section{introduction}

Long-range interactions, such as the dipole-dipole interaction, the Van der Waals forces, etc., play important roles in cold atomic systems and can result in a variety of intriguing physical effects \cite{dauxois2002dynamics,dudin2012strongly,landig2016quantum,mottl2012roton,ritsch2013cold,blass2018quantum,georges2018light,norcia2018cavity,kroeze2018spinor}.
In recent years, photon-mediated long-range interaction between atoms inside an optical cavity has also received wide attentions as these systems provide an opportunity to engineering atom-atom interaction in a highly controllable manner \cite{van2013photon,schutz2014prethermalization,davis2019photon,aron2016photon,welte2018photon,landig2015measuring,mottl2012roton,landig2016quantum},
 where both the range and the strength of the interaction can be tailored \cite{vaidya2018tunable,PhysRevA.82.043612,PhysRevLett.122.193601}. 
For instance, the cavity-mediated long-range spin-spin interaction can be engineered to realize various frustrated models \cite{PhysRevLett.107.277201,gopalakrishnan2009emergent}.  The competition between the short- and long-range interactions induced in cavity also greatly enriches the physics of quantum phase transitions, which is unattainable in other setups \cite{landig2016quantum}. For fermions, such long-range interaction can also result in exotic topological superfluids featuring Majorana fermions \cite{PhysRevLett.123.133601}.

Recently, a hybrid atom-optomechanical system made up of a membrane inside a cavity and cold atoms residing in an optical lattice outside the cavity has attracted wide attentions \cite{vogell2013cavity,bennett2014coherent,vogell2015long,mann2018nonequilibrium,tan2015hybrid,mann2018enhancing,vochezer2018light,jockel2015sympathetic,mann2019tuning}.
This system can not only serve as a platform to explore the coupling between the mechanical modes and other physical systems \cite{vochezer2018light}, but also provide a toolbox to engineer the quantized lattice vibrations \cite{jockel2015sympathetic,christoph2018combined}.
As the cavity and the outside lattice are separate and can be manipulated almost independently,
both the lattice potential and the effective long-range atom-atom interaction induced by the quantized optical modes are highly controllable. For example, in previous studies where the atoms are confined inside the cavity, the possible lattice spacing along the cavity axis is usually determined by the cavity length and the cavity mode functions \cite{colombe2007strong,brennecke2007cavity,nagy2010dicke}. This restriction is no longer present in this hybrid atom-optomechanical system.
For Bose-Einstein condensates, it has been theoretically predicted that the atomic cloud can experience a non-equilibrium quantum phase transition from a localized symmetric state to a shifted spontaneous-symmetry-broken state due to the presence of induced membrane-atom coupling \cite{mann2018nonequilibrium,mann2019tuning}.
Across the transition, the lattice can be either left- or right- shifted depending on the sign of the membrane displacement, which reflects the breaking of the internal Z$_2$ symmetry of the system.
The relevant steady-state many-body phase diagram and non-equilibrium quantum phase transition for spinor system have also been considered \cite{gao2019steady,mann2019tuning,PhysRevA.100.053616}.

In all previous theoretical studies of the atom-optomechanical system, the effect of cavity-mediated global interaction among atoms has been neglected under the assumption that such interaction is very weak. The validity of this assumption, however, is not thoroughly investigated.
Usually, the cavity-mediated effective interaction between atoms can result in various novel self-organized structures \cite{baumann2010dicke,bakhtiari2015nonequilibrium,klinder2015observation} and strongly correlated phases \cite{busche2017contactless,motzoi2018precise}. A careful study of this effect in the hybrid atom-optomechanical system is thus highly desirable. This provides the main motivation of the current work.

In this work, we derive the explicit form of the cavity-mediated effective atom-atom interaction in this hybrid atom-optomechanical system. We show that this effective interaction is qualitatively different from the one when the atoms are inside the cavity \cite{van2013photon,schutz2014prethermalization,davis2019photon,aron2016photon,welte2018photon,landig2015measuring,mottl2012roton,landig2016quantum,vaidya2018tunable,PhysRevA.82.043612,PhysRevLett.122.193601}. In particular, the effective interaction in the current situation is non-uniform and site-dependent.
With this effective interaction taken into account, we consider the steady-state phase diagram of the system in the mean-field level.
We show that the intrinsic Z$_2$-type symmetry of the atom-membrane coupling is explicitly broken by the induced global interaction, where a first-order super-radiation phase transition of mechanical modes is favored for large membrane-atom coupling with the presence of a right-moved lattice order (RLO).
For even stronger effective interaction, the condensate spontaneously breaks into peaks with imbalanced onsite occupations.
These peaks form an approximate lattice-like density-wave order (DWO) with, however, unequal spacing between adjacent peaks.
Meanwhile, the onsite occupation increases monotonously away from the cavity, and the transition from the usual lattice order to the DWO is of first-order.
Finally, the stability and the excitation spectra of relevant phases are also discussed.

The paper is organized as follows. In section II, we present a detailed derivation about the effective Hamiltonian of the model, where the underlying physics about the induced interaction is discussed.
In section III, we introduced the mean-field treatment of the system, where an effective Gross-Pitaevskii (GP) equation of the condensate is obtained.  In section IV, we consider the effects of the induced global  nonlinear interaction, and the properties of the quasi-lattice like mode are discussed in some details.
We provide the phase diagram of the system in section V, where the relevant first-order and second-order phase transitions are also discussed. We conclude the paper in section VI. Much of the technical details can be found in the appendices.

\section{\label{sec:model}model Hamiltonian}
\begin{figure}[htpb]
	\centering
	\includegraphics[width=0.45\textwidth]{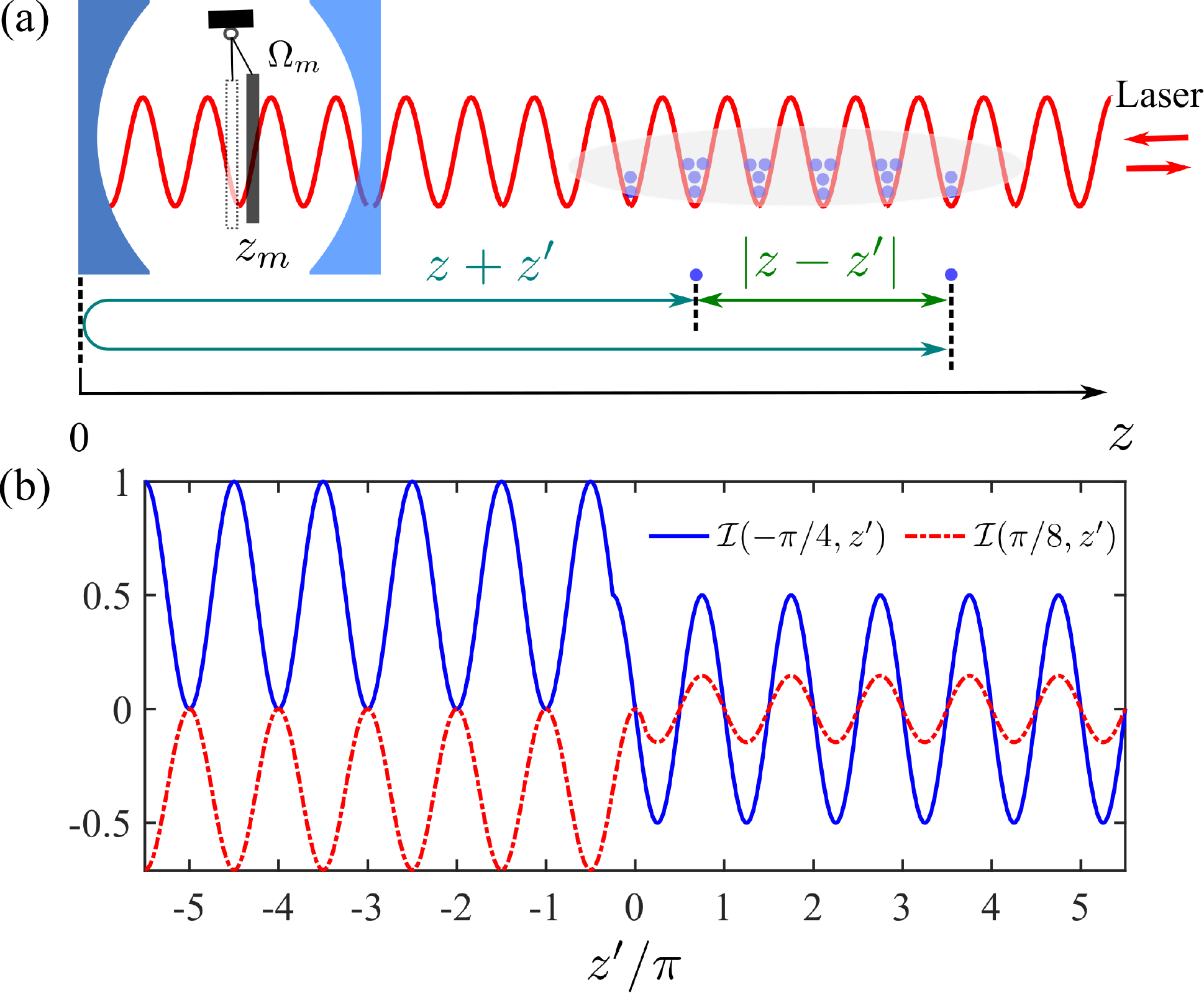}
	\caption{\label{fig:f0}
	(a) Schematic diagram of atom-Optomechanical system. There exist two different paths ($\vert z-z^\prime\vert$, $z+z^\prime$) for intermediated photons. The origion $z=0$ is defined to be the position of the left cavity mirror.
	(b) Effective interaction $\mathcal{I}(z,z^\prime)$ with $z=-\pi/4$ and $z=\pi/8$.
	For convenience, we have shifted the origin of the coordinate to the middle of the lattice.}
\end{figure}

The hybrid atom-optomechanical system we consider here consists of a membrane inside an optical cavity and an ensemble of $^{87}\text{Rb}$ Bose-Einstein condensate outside of the cavity confined in an external optical lattice \cite{vogell2013cavity,vogell2015long,mann2018nonequilibrium,mann2018enhancing,mann2019tuning}, as schematically shown in Fig.~\ref{fig:f0}(a). The lattice potential results from a laser beam propagating towards left along the $z$-axis, and a counter-propagating reflected light beam from the cavity mirror.
The mechanical mode for the membrane can be described as
\bea
\hat{H}_\text{m}=\hbar\Omega_\text{m}\hat{a}^{\dagger}\hat{a}
\eea
with a single mechanical frequency $\Omega_\text{m}$.
The usual many-body Hamiltonian for the condensate can be written as
\bea
	\hat{H}_a=\int dz~\hat{\psi}^{\dagger}(z)\mathcal{H}_0\hat{\psi}(z)+\frac{g}{2}\int dz~\hat{\psi}^{\dagger}\hat{\psi}^{\dagger}\hat{\psi}\hat{\psi}
\eea
with bosonic field operator $\hat{\psi}(z)$, $s$-wave interaction strength $g$, and atom number $\hat{N}=\int dz~\hat{\psi}^{\dagger}\hat{\psi}$. The single-particle Hamiltonian reads \bea	\mathcal{H}_0=-\hbar\omega_R \partial^2/\partial z^2+V\sin^2(z) \eea
with recoil energy $\omega_R=\hbar k^2_l/(2m)$ and amplitude $V$ of the optical lattice, where $m$ stands for the mass of rubidium atom and $k_l$ represents wave number of the laser field. In writing $H_a$ and $\mathcal{H}_0$, we have used the dimensionless coordinates $z$ and set $1/k_l$ as the units for
length.

The condensate couples to the mechanical modes through a broad-band laser modes $\hat{b}_{\omega}$ described as
\bea	\hat{H}_l=\int^{\omega_l+\theta}_{\omega_l-\theta}d\omega~\hbar(\omega-\omega_l)\hat{b}^{\dagger}_\omega \hat{b}_\omega \eea
with its central frequency $\omega_l$ and spectral width $2\theta$.
The bandwidth $2\theta$ of the field modes, namely the line-width of the cavity, should be much larger than the recoil frequency $\omega_R$ and the characteristic frequency $\Omega_m$ of the membrane.
In our case, the laser fields take the form
\bea \hat{b}_{\omega}  \rightarrow \hat{b}_{\omega} + b_l \delta(\omega-\omega_l).  \eea
Physically the laser modes play two distinct roles: 
First, the light mode $\langle \hat{b}_{\omega} \rangle = b_l \delta(\omega-\omega_l)$ has a strong field strength $b_l$ at the central frequency $\omega_l$, which induces the external potential $V_L=V\sin^2(z)$, together with an effective atom-laser coupling
\begin{align}
	\hat{H}_{al}=\lambda_a\int \frac{d\omega}{\sqrt{2\pi}}(\hat{b}_\omega+\hat{b}^{\dagger}_\omega)\int \hat{\psi}^{\dagger}\sin(z)\sin(\frac{\omega}{\omega_l}z)\hat{\psi} dz.
\end{align}
Second, after entering into the cavity, these laser modes also lead to  membrane-light coupling described by the following Hamiltonian
\begin{align}
	\hat{H}_{ml}=\lambda_m(\hat{a}+\hat{a}^{\dagger})\int d\omega~(\hat{b}_\omega+\hat{b}^{\dagger}_\omega)/\sqrt{2\pi}.
\end{align}
Here $\lambda_m$ and $\lambda_a$ are the relevant coupling strengths.

\begin{widetext}

In the case of a broad-band light field and in the bad-cavity limit, we solve the Heisenberg equations of motion for operators $\hat{a}$, $\hat{b}$, and $\hat{\psi}$. After substituting the formal solution of $\hat{b}_{\omega}(t)$ into the equations of motion for $\hat{a}$ and $\hat{\psi}$  (see Appendix \ref{sec:ham} for details), we find
\bea
i\hbar\frac{\partial}{\partial t}\hat{a}&=&\hbar\Omega_m \hat{a}-\Lambda\int dz~\hat{\psi}^{\dagger}(z)\sin(2z)\hat{\psi}(z), \\
i\hbar \frac{\partial}{\partial t} \hat{\psi}(z)&=&\Big\{\mathcal{H}_0+g\hat{\psi}^{\dagger}\hat{\psi}-\Lambda(\hat{a}+\hat{a}^{\dagger})
	\sin(2z)  -\Gamma\int dz^{\prime}\hat{\psi}^{\dagger}(z^{\prime})\hat{\psi}(z^{\prime})\mathcal{I}(z,z^\prime) 
	\Big\}\hat{\psi}(z)
\eea
\end{widetext}
with
\bea
\Lambda&=&\lambda_m\lambda_a/(2\hbar), \hspace{0.5cm} \Gamma = \lambda^2_a/(2\hbar), \\
\mathcal{I}(z,z^\prime) &=& [\sin(z^{\prime}+z)-\sin\vert z^{\prime}-z\vert]\sin(z^{\prime})\sin(z), \label{Izz}
\eea
where  we have omitted the Langevin noise terms for simplicity.
The above equations indicate that the effective membrane-atom coupling Hamiltonian can be written as
\begin{align}
	\hat{H}_{ma}=-\Lambda(\hat{a}+\hat{a}^{\dagger})\int dz \, \hat{\psi}^{\dagger}(z)\sin(2z)\hat{\psi}(z)\,.
\end{align}
In addition, the system also gives rise to an effective cavity-mediated global atom-atom interaction, described by the Hamiltonian
\bea
	\hat{H}_{lr}=\frac{\Gamma}{2}\int dz \int dz^{\prime} \hat{\psi}^{\dagger}(z) \hat{\psi}^{\dagger}(z^{\prime})\mathcal{I}(z,z^\prime)\hat{\psi}(z^{\prime})\hat{\psi}(z).
	\label{eq:lr}
\eea
Thus the total effective Hamiltonian, after eliminating the laser modes, only contains the membrane and the atomic degrees of freedom and reads
\begin{align}
	\label{eq:ham}
	\hat{H}_{\text{eff}}=\hat{H}_{m}+\hat{H}_{a}+\hat{H}_{ma}+\hat{H}_{lr}.
\end{align}
Equation (\ref{eq:lr}), along with (\ref{Izz}), represents one of the main results of the work.
Physically, since all the atoms are coupled to the same laser fields $\hat{b}_{\omega}(t)$, these quantized modes can thus be used as a bus for mediating the long-range interaction between atoms.
Here, the two atoms located at $z$ and $z'$ can be linked by the intermediating fields $\hat{b}_{\omega}(t)$ through two different paths, as shown in Fig. \ref{fig:f0}(a).
The first path corresponds the shortest distance $|z-z'|$ between the two atoms.
In the second path, after leaving the first atom at $z$, the intermediating photon is reflected back by the cavity mirror before it reaches the second atoms located at $z'$.
The total distance traced by the photon is therefore $z+z^\prime$.
This explains the origin of the two different sinusoidally modulated interaction terms contained in the
effective global interaction $\hat{H}_{lr}$.

In the absence of the effective interaction $\hat{H}_{lr}$, it has been shown theoretically that the atoms experience a second-order phase transition from a localized symmetric state with $X_m = \avg(a+a^{\dag})=0$ to a shifted symmetry-broken state with $X_m \neq 0$ as the membrane-atom coupling $\Lambda$ increases \cite{mann2018nonequilibrium,mann2018enhancing,mann2019tuning,gao2019steady}.
Compared with the usual case with the atoms inside the cavity, here the lattice spacing is not changed before and after the transition point \cite{colombe2007strong,brennecke2007cavity,nagy2010dicke}.

The induced global interaction appeared in $\hat{H}_{lr}$ exhibits interesting features.
Specifically, if we focus on the atom fixed at $z$, the effective interaction $\mathcal{I}(z,z^\prime)$ reduces to
\bea
\mathcal{I}(z,z^\prime) = \left\{
                            \begin{array}{ll}
                              -\sin 2z' \sin^2 z, & \mbox{ for } z'>z; \\
                              -\sin 2z \sin^2 z', & \mbox{ for } z'<z.
                            \end{array}
                          \right.
\eea
Therefore, when $z \neq j \pi$, the effective long-range interaction between atoms at $z$ and $z'$ shows different site-dependent features for $z'>z$ and $z'<z$. In
Fig.~\ref{fig:f0}(b), we plot the effective interaction $\mathcal{I}(z,z^\prime)$ for fixed $z=-\pi/4$ and $z=\pi/8$.
One can see that when $z=j\pi+\delta z$ is slightly displaced from the local minima $j\pi$ of the lattice potential $V_L$, the mean effective interaction $\mathcal{I}(z,z^\prime)$ for $z'<z$ takes positive and negative values depending on the sign of $\delta z$.
We stress that this site-dependent feature of the induced atom-atom interaction is very different from those obtained for atoms inside the cavity \cite{PhysRevLett.107.277201,gopalakrishnan2009emergent}, where the interaction usually only depends on $|z-z'|$.
The induced global non-uniform interaction can affect the steady state of the system significantly and lead to unexpected physics, which we will focus in the following.

\section{Mean-field approximation}

For condensate with large atomic number $N$ and neglectable quantum fluctuations, we can employ the mean-field approximation, and replace the operators $\hat{\psi}(z)$ and $\hat{a}$ with their mean values. After making substitutions $\hat{\psi}(z)\rightarrow \sqrt{N}\varphi(z)$ and $\hat{a}\rightarrow\sqrt{N}\alpha$, the Heisenberg equations of motion for operators $\hat{\psi}(z)$ and $\hat{a}$ can then be rewritten as
\bea
	i\hbar\partial_t \alpha &=& \hbar(\Omega_m-i\gamma)\alpha-\Lambda\sqrt{N}\int dz |\varphi|^2\sin(2z),  \label{alpha}\\
	i\hbar\partial_t\varphi(z) &=& \Big\{\mathcal{H}_0 -\Lambda\sqrt{N}(\alpha+\alpha^{\dag})\sin(2z)  \notag \\
	&& \hspace{.9cm} +gN\vert\varphi(z)\vert^2 +\Gamma N\chi[\varphi,z] \Big\}\varphi(z) ,  \label{eq:gp1}
\eea
with the functional
\bea
\chi[\varphi,z]&=&\ \int \!\! dz^{\prime}~\vert\varphi(z^{\prime})\vert^2\mathcal{I}(z,z^\prime).
\eea
In Eq.~(\ref{alpha}), we have introduced a damping rate $\gamma$ for the mechanical mode, and the normalization condition for $\varphi$ reads $\int dz~\vert\varphi(z)\vert^2=1$.

To simplify the discussion, we further assume that the membrane reaches its steady state very quickly due to its fast damping rate, and hence we can take $\partial_t\alpha=0$. This assumption gives
\begin{align}
	\alpha=\frac{\Lambda\sqrt{N}}{\hbar(\Omega_m-i\gamma)}\kappa[\varphi],
\end{align}
with the functional
\bea
\kappa[\varphi]=\int dz'~\vert\varphi(z')\vert^2\sin(2z').
\eea
After substituting this back to Eq. (\ref{eq:gp1}), we arrive at the effective GP equation for the condensate
\begin{align}
	\label{eq:gp}
	i\hbar\partial_t\varphi(z)=\Big\{&\mathcal{H}_0-\tilde{\Lambda}\kappa[\varphi]\sin(2z) +\tilde{g}\vert\varphi(z)\vert^2 \notag \\
	&+\tilde{\Gamma}\chi[\varphi,z]\Big\}\varphi(z),
\end{align}
with the following interaction parameters
\begin{align}
	\tilde{\Lambda}&=N\beta\Lambda^2/(\hbar\Omega_m),&&  \!\!\!  \beta=2\Omega^2_m/(\Omega^2_m+\gamma^2), \nn \\
	\tilde{g}&=Ng, && \!\!\!  \tilde{\Gamma}=N \Gamma. \nn
\end{align}
The relative strength between the induced long-range interaction $\tilde{\Gamma}$ and the effective atom-membrane coupling $\tilde{\Lambda}$ can then be determined by $\lambda_m$, $\Omega_m$, and $\gamma$ etc.
In appendix B, we have provided an explicit estimation of these parameters based on current experimental conditions, which also covers the parameter ranges discussed in the following.

We solve Eq.~(\ref{eq:gp}) using the imaginary-time evolution method to obtain the ground state.
For a deep lattice potential $V_L=V\sin^2(z)$ with $V\gg \{\tilde{g},\tilde{\Lambda},\tilde{\Gamma}\}$, the atoms mainly accumulate around its local minima at $z^0_j = j\pi$ and thus form a lattice order.
The presence of the effective membrane-atom coupling $\hat{H}_{ma}$ introduces an additional potential proportional to $V_{ma}=-\Lambda X_m\sin(2z)$. 
This additional potential $V_{ma}$ shares the same period as $V_L \propto \cos(2z)$, but features a relative phase shift.
When the membrane-atoms coupling is weak, $V_L$ dominates and the aforementioned lattice order remains unchanged.
However, for sufficiently large $\Lambda$, $V_{ma}$ can drive the lattice to move to left or to right depending on the sign of the membrane displacement $X_m$.
The right- and left-moved lattice orders are degenerate when the effective atom-atom interaction is absent, i.e., $\hat{H}_{lr}=0$.
Therefore, a second order phase transition takes place in this process accompanied with a spontaneous breaking of the Z$_2$ symmetry.
We stress that, for atoms inside the cavity, similar super-radiant phase transition has also been predicted and observed, which usually accompanies with a change of the periodicity of the lattice before and after the transitions.
In this hybrid system, by contrast, the lattice period can remain unchanged when the transition occurs.

\section{\label{sec:eff}Effects of the global non-uniform interaction}

When $\hat{H}_{lr} \neq 0$, the presence of the global atom-atom interaction can result in many novel features, which will be the focus of  this section.

\begin{figure}[htpb]
	\centering
	\includegraphics[width=0.45\textwidth]{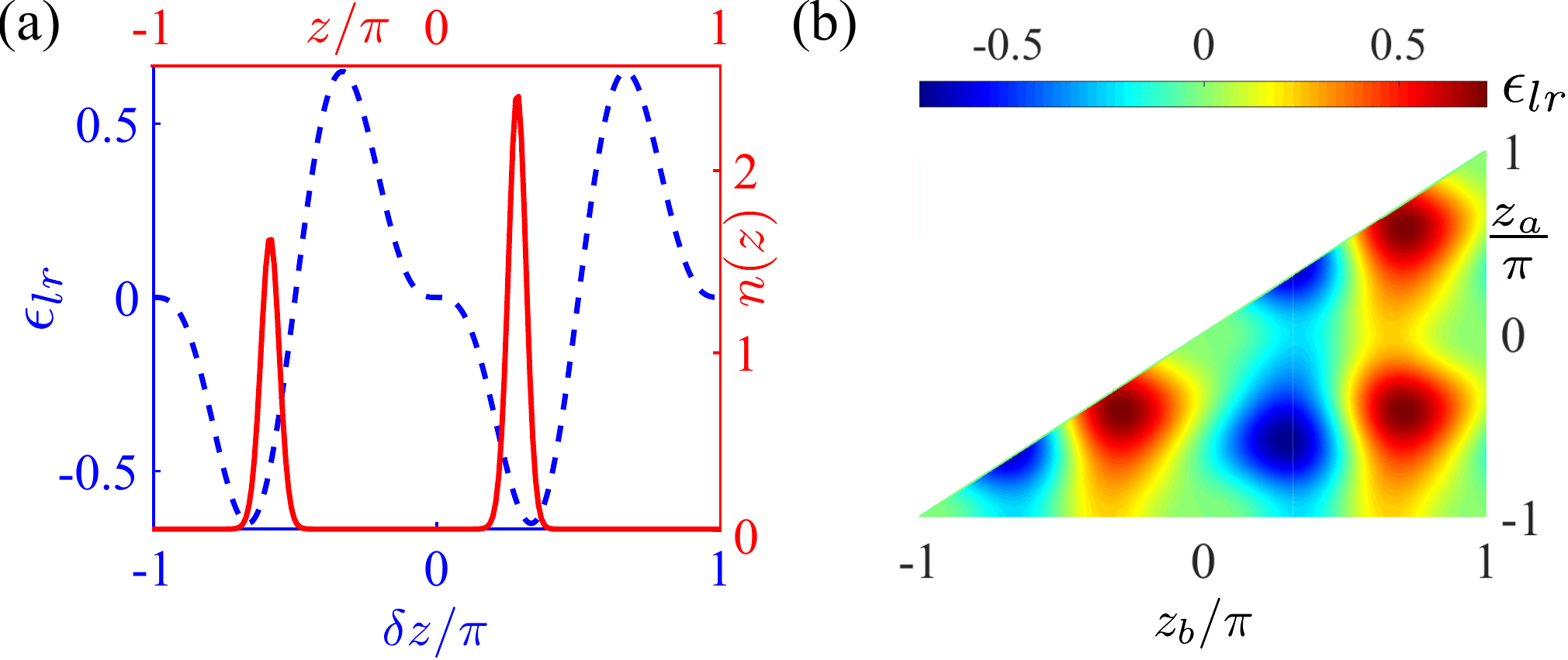}
	\caption{\label{fig:lrint}
	(a) Contour plot of $\epsilon_{lr}$ for two-sites case in $z_a-z_b$ plane with $z_a<z_b$, $n_a=0.4$ and $n_b=0.6$.  The minimal point is $(z_a,z_b)\approx(-0.58,0.29)\pi$.
		(b) Density distribution $n(z)$ (red line) of atoms due to the presence of long-range interaction $H_{lr}$. Here we assume the lattice is not presented ($V=0$). Other parameters read: $g=0$, $N=10^4$, $m=\hbar=\omega_R=1$, $\Omega_m=100$, $\gamma=10$, $\tilde{\Lambda}=0$ and $\tilde{\Gamma}=625$. The two density peaks locates around $z_a\approx-0.58\pi$ and $z_b\approx0.29\pi$ with $n_a/n_b\approx2/3$.
	The dashed blue line shows the odd function $\epsilon_{lr}=-\sin^2(\delta z)\sin(2\delta z)$.	
For convenience, we have shifted the origin of the coordinate to the middle of the lattice.
	}
\end{figure}

First, we note that  the nonlinear interaction does not preserve the Z$_2$ symmetry.
Since an arbitrarily weak long-range interaction can break the Z$_2$ symmetry of the system, the lattice favors to move once the coupling $\Lambda$ surpasses the transition point $\Lambda_c$.
To show this, we consider a simplified wavefunction for the condensate in the deep lattice limit as
\bea
\varphi(z) = \sum_{j=1}^L c_j |z=z_j\rangle,  \label{eq:LsiteWF}
\eea
where $\sum_j |c_j|^2=1$ and $z_j$ represents the location of the $j$-th wave-packet. The basis $|z\rangle$ satisfies $\langle z |z'\rangle =\delta(z-z')$. We also assume $z_j = j\pi +\delta z$ with $\delta z$ the overall shift of the lattice order.
A simple algebra shows that the mean interaction energy for an $L$-site lattice reads
\bea
E_{lr} = \avg(\hat{H}_{lr})/L = \frac{\tilde{\Gamma}}{2}\epsilon_{lr} \label{eq:interactionEn}
\eea
with
\bea
\epsilon_{lr}=-\sin^2 (\delta z) \sin(2\delta z),
\eea
which is an odd function of $\delta z$ and reaches the minimum value at $\delta z =\pi/3$, as depicted in Fig.~\ref{fig:lrint}(a).
Therefore, a right-moved lattice is always favored which breaks the intrinsic Z$_2$ symmetry of the original model.
In addition, such weak long-range interaction also makes the phase transition to be of first order (see analysis in Appendix \ref{sec:weaklong}).

Second, for stronger interaction strength $\Gamma$, the nonlinear interaction can induce an effective lattice potential, which can change both the density distribution of the condensate and the lattice pattern of the ground state.
In the case of very strong long-range interaction, the original periodic lattice pattern of the condensate induced by $V_L$ becomes unstable.
The system supports a series of isolated Gaussian wave-packets.
These isolated packets exist even when the lattice potential $V_L \propto \sin^2(z)$ become negligible compared with the nonlinear interactions.
The spacings between adjacent packets are not constant. Hence we call this a quasi lattice-like pattern.
In addition, the peak values of these wave-packets are also not uniform and increase as their distance away from the cavity increases.
The presence of the self-adapted lattice-like density wave order (DWO) represents another key feature induced by the effective global nonlinear interaction.

We stress that the presence of such quasi lattice-like pattern
can be completely attributed to the interaction $H_{lr}$, as this pattern exists even when the lattice trap is absent $V=0$.
Physically, the lattice potential can be tuned by introducing another laser which is slightly misaligned with the former one and generates a lattice with the same lattice spacing.
To present a simple picture of the emergence of the DWO order,  let us consider the simplest case with two Gaussian wave-packets localized within the regime \bea -\pi\leq z_a <z_b \leq \pi.\eea
The condensate wavefunction reads
\bea \varphi(z)=c_a|z=z_a\rangle +c_b|z=z_b\rangle \eea
with $|c_a|^2+|c_b|^2=1$.
The inter-site part of the interaction can be simplified as
\bea  E^{int}_{lr}=-\tilde{\Gamma} n_an_b\sin^2(z_a)\sin(2z_b)  \eea with $n_{a,b} = |c_{a,b}|^2$.
It is easy to check that to minimize the interaction energy $E_{lr}$ shown in Eq. (\ref{eq:interactionEn}), we should set $z^0_a=-2\pi/3$ and $z^0_b=\pi/3$, as shown in Fig.~\ref{fig:lrint}(a).
Around $(z_a,z_b)=(-2\pi/3,\pi/3)$, we have
\bea
\frac{\partial E^{int}_{lr}}{\partial z_a} |_{(z_a,z_b)=(-\frac{2\pi}{3},\frac{\pi}{3})} &=& -\frac{3}{4}\tilde{\Gamma}n_an_b <0, \\
\frac{\partial E^{int}_{lr}}{\partial z_b}|_{(z_a,z_b)=(-\frac{2\pi}{3},\frac{\pi}{3})} &=& \frac{3}{4}\tilde{\Gamma} n_an_b >0.
\eea
Therefore, the interaction $E^{int}_{lr}$ can be further reduced if we choose a modified configuration with
\bea -2\pi/3 < z_a' < z_b' < \pi/3 \eea such that $z_b'-z_a'<\pi$.
Similar analysis also indicates that $n_a<n_b$ is favored to obtain an overall lower energy
\bea
E_{lr} &=& -\frac{\tilde{\Gamma}}{2} \Big[ n_a^2\sin^2(z_a)\sin(2z_b) + n_b^2\sin^2(z_b)\sin(2z_a)  \Big] \nn \\&& +  E^{int}_{lr}.
\eea
This is also verified numerically, where $E_{lr}$ is minimized when $n_a \simeq 0.4$, $n_b \simeq 0.6$, $z_a\approx-0.58\pi$ and $z_b\approx0.29\pi$, as shown in Fig.~\ref{fig:lrint}(b).



The above discussion can also be generalized to $L$-site case.
As in the two-site case discussed above, the calculation indicates that these wave-packets tend to be localized at positions with intervals less than $\pi$. Furthermore, the spacings between adjacent wave-packets are not constant.
To illustrate this, we consider the simplified condensate wavefunction shown in Eq. (\ref{eq:LsiteWF}).
The corresponding interaction energy can be written as
\bea
	\label{eq:lrinter}
	E_{lr}= E^-_{lr} + E^+_{lr}
\eea
with
\bea
E^-_{lr} &=& \frac{\tilde{\Gamma}}{2}\sum_{j=1}^{L}n_j\sin^2(z_j)\Big[\sum_{i<j}n_i\sin(2z_j) -\sum_{i>j}^Ln_i\sin(2z_j) \Big],  \nn \\
E^+_{lr} &=& -\frac{\tilde{\Gamma}}{2}\sum_{j=1}^{L}n_j\sin^2(z_j) \sum_{i=1}^{L}n_i\sin^2(z_i). \nn
\eea
where we have set $n_j=\vert c_j\vert^2$, and $E^-_{lr}$ and $E^+_{lr}$ correspond to two different terms in Eq. (\ref{eq:lr}) depending on $\sin(|z-z'|)$  and $\sin(z+z')$ respectively.
This leads to
\bea
E_{lr} = -\tilde{\Gamma}\sum_{j=1}^{L}n_j\sin^2(z_j)\Big[\frac{1}{2}n_j\sin(2z_j) +  \sum_{k>j}^{L}n_k\sin(2z_k)\Big]. \nn \\
\eea
The first term corresponds to on-site interaction which is minimized when $z_j = j\pi+\pi/3$ with the lattice interval $\Delta  =\pi$.
The last term describes the long-range interaction between different sites, and depends closely on the index order $j$ along the $z$-axis.
Therefore, the effective potential at position $z_j$ due to $E_{lr}$ reads
\bea
	\mathcal{V}(z_j)&=&\frac{\partial}{\partial n_j}E_{lr}=-\tilde{\Gamma}\Big[\sin^2(z_j)\sum_{k=j}^{L}n_k\sin(2z_k) \nn \\
	&& +\sum_{k=1}^{j-1}n_k\sin^2(z_k)\sin(2z_j)\Big].  \label{eq:vj}
\eea

\begin{figure}[htpb]
	\centering
	\includegraphics[width=0.45\textwidth]{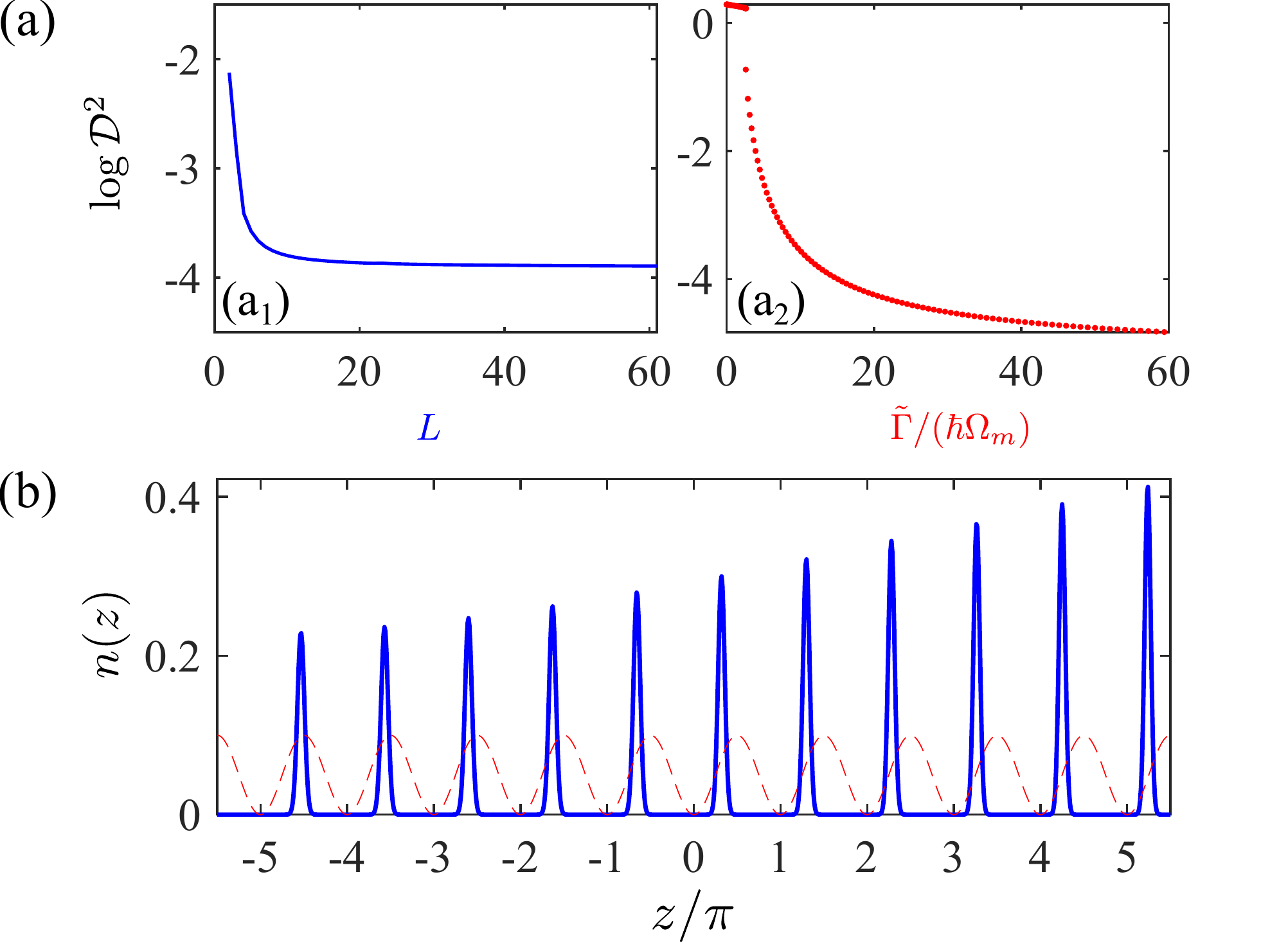}
	\caption{\label{fig:wfdev}
	(a) Variance $\mathcal{D}^2$ as a function of the total lattice site $L$ with $\tilde{\Gamma}/(\hbar\Omega_m)=12.5$ (blue line) and $\tilde{\Gamma}/(\hbar\Omega_m)$ with $L=11$ (red line).
	(b) Numerically obtained density distributions $n(z)$ with $L=11$, $\tilde{\Gamma}/(\hbar\Omega_m)=12.5$. Red line represents the scaled optical lattice.
	In both figures we set $\omega_R=1$, $V=200$, $\Omega_m=100$, $\gamma=10$, $\tilde{g}=10$, $N=10^4$, $m=\hbar=1$, $\tilde{\Lambda}/(\hbar\Omega_m)=0.495$.
	For convenience, we have shifted the origin of the coordinate to the middle of the lattice.
	}
\end{figure}

\begin{figure}[htpb]
	\centering
	\includegraphics[width=0.45\textwidth]{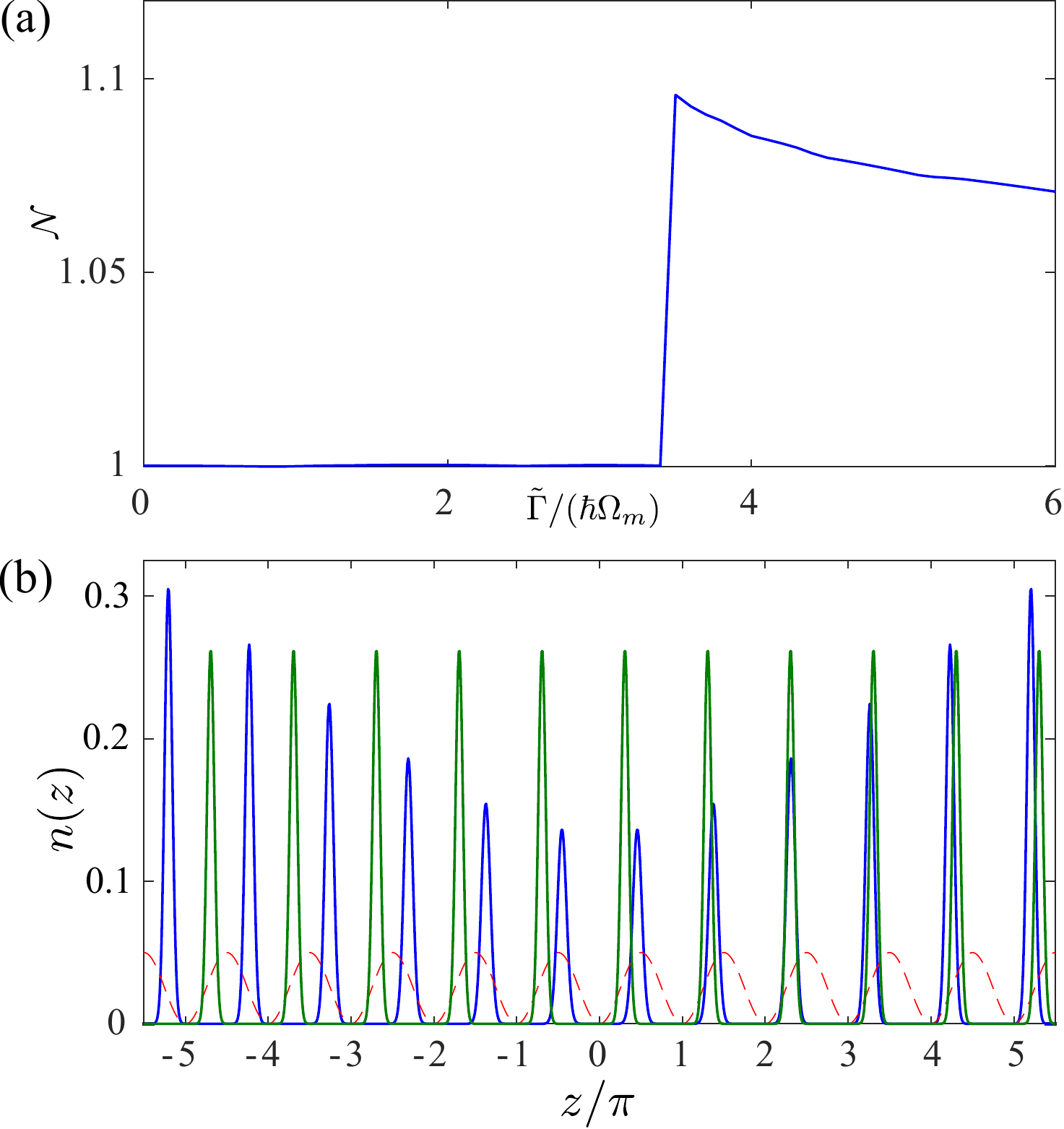}
	\caption{\label{fig:zn}
	(a) Population imbalance $\mathcal{N}$ along with interaction strength $\tilde{\Gamma}$ in case of $\tilde{\Lambda}/(\hbar\Omega_m)=0.1$.
	(b) Density distributions $n(z)$ in case of only $E^-_{lr}$ (blue line) and only $E^+_{lr}$ (green line) with $\tilde{\Lambda}/(\hbar\Omega_m)=0.1$, $\tilde{\Gamma}/(\hbar\Omega_m)=5$. Red line represents the scaled optical lattice.
	For convenience, we have shifted the origin of the coordinate to the middle of the lattice.
	We set other parameters as: $\omega_R=1$, $V=200$, $\Omega_m=100$, $\gamma=10$, $\tilde{g}=10$, $N=10^4$, $m=\hbar=1$, $L=11$.
	}
\end{figure}

Usually, the interaction energy $E_{lr}$ is minimized when the effective potential $\mathcal{V}(z_j)$ is also minimized as far as possible.
Using this simple observation, we can then estimate the mean distance of these wave-packets.
For the leftmost wave-packet, we have
\bea
\mathcal{V}(z_1)\propto-2\sin^2(z_1)\sum_{j=1}^{L}n_j\sin(2z_j)
\eea
and for the rightmost one, we have
\bea
\mathcal{V}(z_L)\propto-2\sin(2z_L)\sum_{j=1}^{L}n_j\sin^2(z_j).
\eea
It is easy to check that these two potentials reach their respective minimum  when \blue{$z_1=\bar{z}_1$ and $z_L=\bar{z}_L$ where
\bea \sin^2(\bar{z}_1)=1  \mbox{, and } \sin(2\bar{z}_L)=1. \eea
}
Here, without loss of generality, we assume $z_j\ge0$ for all $j \in \{1,\cdots,L \}$.
In this case, we have
\bea \bar{z}_1=\frac{\pi}{2} \mbox{, and } \bar{z}_L=(L-1)\pi+\frac{\pi}{4}.\eea
Therefore, if these $L$ wave-packets are equally spaced with the shortened interval
\begin{align}
	\label{eq:delta}
	\bar{\Delta}=\frac{\bar{z}_L-\bar{z}_1}{L-1}=\pi-\frac{\pi}{4(L-1)} < \pi,
\end{align}
then the position of the $j$-th wave-packet is estimated as
\bea
\bar{z}_{j}=\bar{z}_1+(j-1)\bar{\Delta}.
\eea

The above analysis is also verified numerically using imaginary-time evolution method. Fig. $\ref{fig:wfdev}$(a) shows
the variance of the estimated $\bar{z}_{j}$ with respective to the exact $z_j$ of the $j$-th wave-packet as
\bea
\mathcal{D}^2 = \frac{1}{L-1}\sum_{j=2}^L|\bar{z}_j-z_j|^2.
\eea
Here the numerically obtained $z_j$ is defined as
\bea
z_j=\int_{z_j}z|\varphi(z)|^2dz/\int_{z_j} |\varphi(z)|^2 dz
\eea
and the integration is performed around the $j$-th Gaussian wave-packet (see Eq. (\ref{eq:gaussian})).
The result shows that $\mathcal{D}^2$ tends to zero quickly for stronger interaction strength $\tilde{\Gamma}$ and longer lattice site $L$.


We also stress that the occupation number $n_j$ is site-dependent, and increases monotonically along with $z_j$.
This is evident if we turn the summation in Eq. (\ref{eq:vj}) into an integral in the limit $L\rightarrow \infty$. A simple algebra gives (see Appendix \ref{sec:effpotential} for details)
\bea
\mathcal{V}(z_j)&\sim& -\frac{\tilde{\Gamma}}{2\pi}(\cos\eta_j+\eta_j\sin\eta_j),
\eea
with $\eta_j = (j-1)\pi/[2(L-1)]$.
Since $\mathcal{V}(z_j)$ decreases as $z_j$ increases, in order to obtain a lower interaction energy $E_{lr}$, the occupation number also increase away from the cavity,  as numerically verified in Fig. $\ref{fig:wfdev}$(b).

To show the varied density of these sites, we introduce the population imbalance  $\mathcal{N}$ defined as
\begin{align}
			\mathcal{N}=\frac{1}{L-1}\sum_{j=2}^L |c_j|^2/|c_{j-1}|^2. 
\end{align}
which quantifies the mean population difference between adjacent sites. Fig. \ref{fig:zn}(a) shows the population imbalance $\mathcal{N}$ as a function of the interaction strength $\tilde{\Gamma}$ with all other parameters fixed. When the induced interaction becomes dominant at large $\tilde{\Gamma}$, $\mathcal{N}$ becomes larger than $1$, which indicates that the occupation number $n_j$ increases at points farther away from the cavity.

We note that the site-dependent feature of $n_j$ can be understood as the competition between the two interaction terms $E^-_{lr}$ and $E^+_{lr}$.
In Fig. \ref{fig:zn}(b), we have also plotted $n_j$ as the function of the lattice site $z_j$ when only the long-range interaction $E^-_{lr}$ (or $E^+_{lr}$) is considered.
The result indicates that the occupation $n_j$ favors an approximated central symmetric pattern with modified lattice spacing for $E^-_{lr}$.
When only  $E^+_{lr}$ is involved, the lattice pattern of the condensate exhibits an overall shift without changing the spacing $\bar{\Delta} = \pi$.
It is the competition of these two different mechanisms that leads to the unique distribution of the $n_j$ in this hybrid system.
We also note that for condensate inside the cavity, both the positions of the sites and the period of the lattice are fixed by cavity parameters and mode functions.
Therefore, the quasi-lattice like order with unequal lattice spacing cannot be supported.

\section{\label{sec:phase}phase diagram}

\begin{figure}[htpb]
	\centering
	\includegraphics[width=0.48\textwidth]{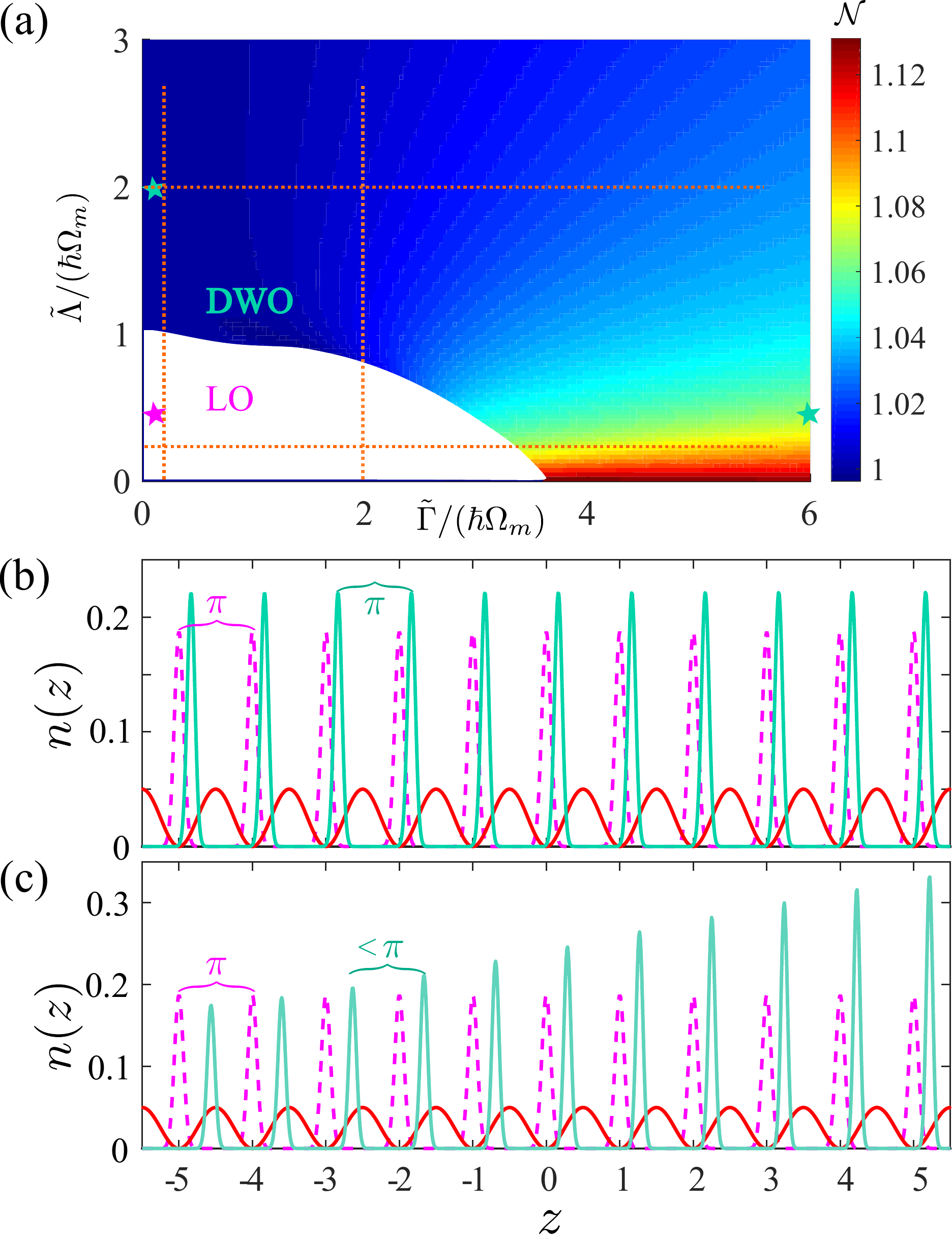}
	\caption{\label{fig:pd}
	(a). Phase diagram in $\tilde{\Gamma}-\tilde{\Lambda}$ plane. ``LO", ``DWO" represent ``Lattice Order", ``Density-Wave Order" respectively. The transition between LO phase and DWO phase is of first-order.
	(b), (c). Representative density distributions at points in the phase diagram marked by five-pointed stars.
	Here red lines are the scaled optical lattices. The dash lines represent the density of a LO state with parameters $\tilde{\Lambda}=0.5\hbar\Omega_m$, $\tilde{\Gamma}=0.1\hbar\Omega_m$. The two representative densities of DWO states are plotted with $\tilde{\Lambda}=2\hbar\Omega_m$, $\tilde{\Gamma}=0.1\hbar\Omega_m$ (b), and $\tilde{\Lambda}=0.5\hbar\Omega_m$, $\tilde{\Gamma}=6\hbar\Omega_m$ (c) respectively.	
	Other parameters are set as: $L=11$, $m=\hbar=1$, $\omega_R=1$, $V=200$, $\Omega_m=100$, $\gamma=10$, $\tilde{g}=10$ and $N=10^4$.
In (b) and (c), we have shifted the origin of the coordinate to the middle of the lattice.
	}
\end{figure}

Based on above discussions, we are now ready to discuss the phase diagram of the system.
For general $\tilde{\Lambda}$ and $\tilde{\Gamma}$, the system supports various lattice patterns.
In the deep lattice limit, these patterns can be described using the variational  wavefunctions
\begin{align}
	\varphi(z)=\sum_{j}c_j\psi_g(z,z_j,\sigma),~\sum_{j}\vert c_j\vert^2=1  \label{eq:gaussian}
\end{align}
with $z_j$ the center of each wave-packet and the Gaussian function reads \bea
\psi_g(z,z_j,\sigma)=\Big(\frac{1}{\pi\sigma^2}\Big)^{1/4}\exp \Big[-\frac{(z-z_j)^2}{2\sigma^2}\Big],
\eea
where parameters $c_j$, $z_j$ and $\sigma$ are determined by minimizing total energy corresponding to this wavefunction.

\begin{figure}[htpb]
	\centering
	\includegraphics[width=0.5\textwidth]{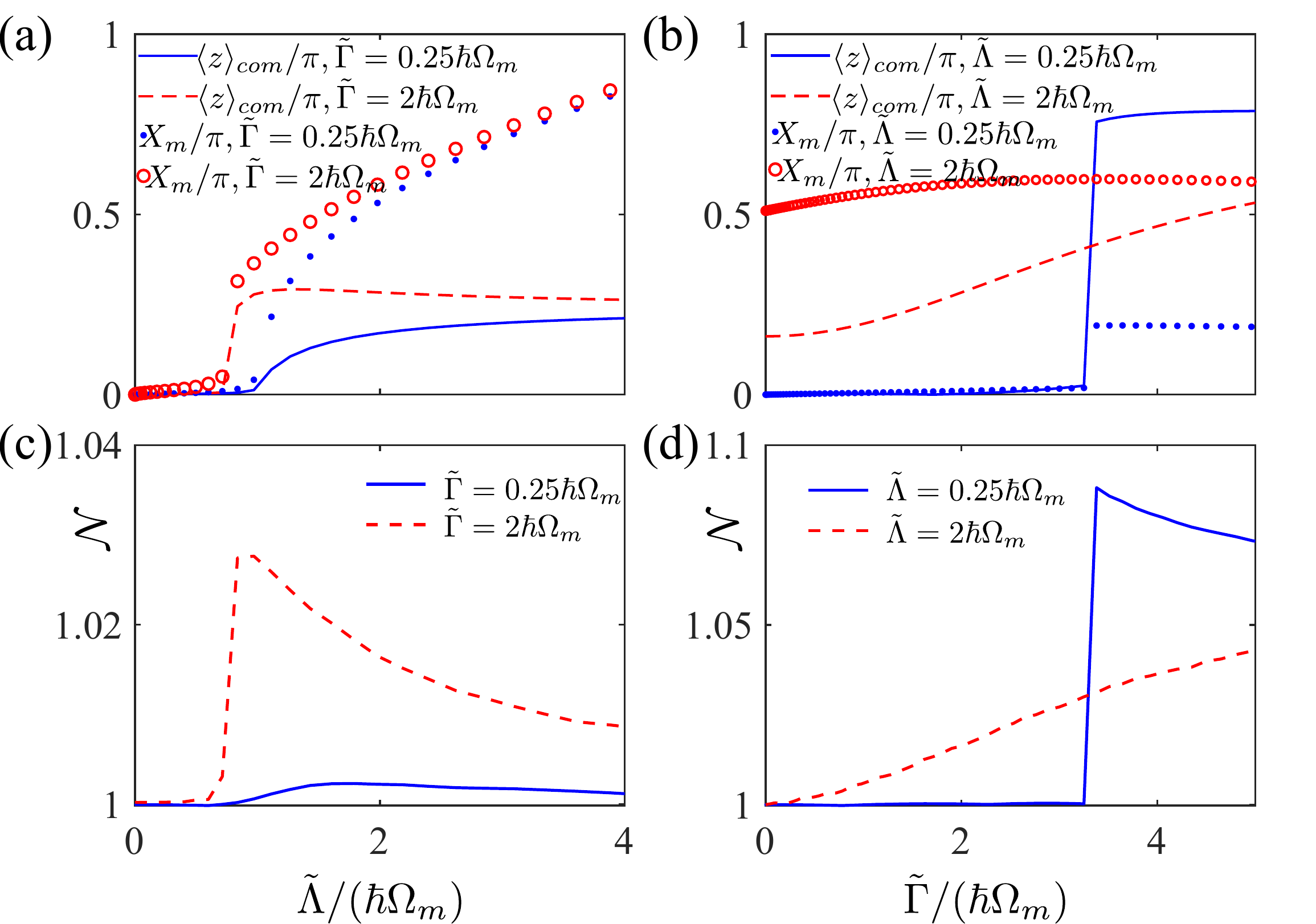}
	\caption{\label{fig:order}
	Order parameters $\langle z\rangle_{com}$, $X_m$ (Fig.a, Fig.b), and $\mathcal{N}$ (Fig.c, Fig.d) along the dashed line shown in figure (5a) as functions of the interaction strength $\tilde{\Lambda}$ and $\tilde{\Gamma}$.
	In all figures we set other parameters as:  $L=11$, $m=\hbar=1$, $\omega_R=1$, $V=200$, $\Omega_m=100$, $\gamma=10$, $\tilde{g}=10$ and $N=10^4$.}
\end{figure}

Figure \ref{fig:pd} shows the obtained phase diagram in the $\tilde{\Gamma}$-$\tilde{\Lambda}$ plane using the imaginary-time evolution method. The result is also checked and confirmed using the variational wavefunctions.
The phase diagram shows novel features which are summarized in the following.

In the absence of the induced global interaction $\tilde{\Gamma}=0$, the system possesses Z$_2$ symmetry. Membrane-atom coupling gives rise to a second-order phase transition from LO to the left- or right-moved LO when $\tilde{\Lambda}$ exceeds the critical value $\tilde{\Lambda}_c$.

The presence of finite $\tilde{\Gamma} \neq 0$ breaks the Z$_2$ symmetry.
Our calculation shows that the critical $\tilde{\Lambda}_c$ decreases monotonously and eventually reaches $0$ as we increase the interaction strength $\tilde{\Gamma}/(\hbar\Omega_m)$.
To show the influence of the global interaction on the transitions, in Fig.~\ref{fig:order}(a) and \ref{fig:order}(c), we plot the order parameters
$\langle z\rangle_{com}$, $\avg(X_m)$, and $\mathcal{N}$ as functions of $\tilde{\Lambda}/\hbar\omega_m$ for fixed $\tilde{\Gamma}=0.25 \hbar\Omega_m$ and $2.0\hbar\Omega_m$ respectively. Here $\langle z\rangle_{com}$ is defined as the overall center-of-mass shift of the condensate.
The calculation shows that all of these parameters jump discontinuously around $\tilde{\Lambda}=\tilde{\Lambda}_c$.
Especially, for $\tilde{\Gamma}/(\hbar\Omega_m) \sim 1$, these jumps becomes more apparent, as shown in Fig.~\ref{fig:order}(a) and \ref{fig:order}(c).
This observation indicates that the transition between the LO phase and the quasi-lattice like DWO is of first order, which is very different from the former case with $\tilde{\Gamma}=0$.
In the DWO regime,  the condensate shows an occupation imbalance $\mathcal{N}>1$ when $\tilde{\Lambda}>\tilde{\Lambda}_c$.
In addition,  $\mathcal{N}$ also becomes smaller for stronger interaction strength $\tilde{\Lambda}$ as the effect of the induced  nonlinear interaction becomes smaller comparatively in this case.

We stress that the presence of the induced nonlinear interaction leads to the change in the order of the transition from LO phase to DWO phase.
To make this point more clear, we assume that when the interaction $\tilde{\Gamma}/(\hbar\Omega_m) \ll 1$ is weak, the periodicity of the system still holds.
In this case, the approximate density distribution of the wavefunction reads
\bea
|\varphi(z)|^2 \simeq \sum_{j=1}^L |c_j|^2\delta(z-z_j)
\eea
with $\vert c_j\vert=1/\sqrt{L}$, and $z_j=j\pi+\delta z$, and its corresponding energy functional is given by (see Appendix \ref{sec:weaklong} for details)
\begin{align}
	E(\delta z)\sim V\sin^2(\delta z)-\frac{\tilde{\Lambda}}{2}\sin^2(2\delta z)-\frac{\tilde{\Gamma}}{2}\sin^2(\delta z)\sin(2\delta z).
\end{align}
Around the phase boundary $\tilde{\Lambda}=\tilde{\Lambda}_c$, the overall shift satisfies $\delta z\sim0$ and we have
\begin{align}
	E(\delta z)=p\delta z^2-\tilde{\Gamma}\delta z^3+q\delta z^4+O(\delta z^5),
\end{align}
with
\bea
p=V-2\tilde{\Lambda}, \quad q=\frac{8\tilde{\Lambda}-V}{3}.
\eea
For weak interaction $\tilde{\Gamma}/(\hbar\Omega_m) \ll 1$, the overall center-of-mass shift $\langle z\rangle_{com}$ of the condensate jumps from $0$ to $\delta z$ after $p$ sweeps across the critical point $p=0$ and can be estimated as \bea \langle z\rangle_{com}=\delta z=3\tilde{\Gamma}/4q.\eea
Since $\delta z >0$, this corresponds to a right-moved lattice order (RLO).
Therefore, within this mean-field treatment, the relevant phase transition is of first-order.

The above transition between different phases are also verified by considering $\langle z\rangle_{com}$, $\avg(X_m)$, and $\mathcal{N}$ as functions of $\tilde{\Gamma}/\hbar\omega_m$ for fixed $\tilde{\Lambda}=0.25 \hbar\Omega_m$ and $2.0\hbar\Omega_m$ respectively, as shown in Fig.~\ref{fig:order}(b) and (d).
For small $\tilde{\Lambda}=0.25 \hbar\Omega_m$, the system supports the LO state  until the global nonlinear interaction $\tilde{\Gamma}$ increases and surpasses a critical value $\tilde{\Gamma}_c$, where the  lattice-like state is favored with nonzero $\langle z\rangle_{com}$, $\avg(X_m)$, and imbalanced on-site occupations $\mathcal{N}$.
For larger $\tilde{\Lambda}=2.0 \hbar\Omega_m > \tilde{\Lambda}^c$, the calculation indicates that the initial RLO states at $\tilde{\Gamma} \ll 1$ changes continuously towards the lattice-like DWO states when $\tilde{\Gamma}$ increases, and the occupation imbalance $\mathcal{N}$ also increases gradually, as shown in Fig.~\ref{fig:order}(d).

\begin{figure}[htpb]
	\centering
	\includegraphics[width=0.45\textwidth]{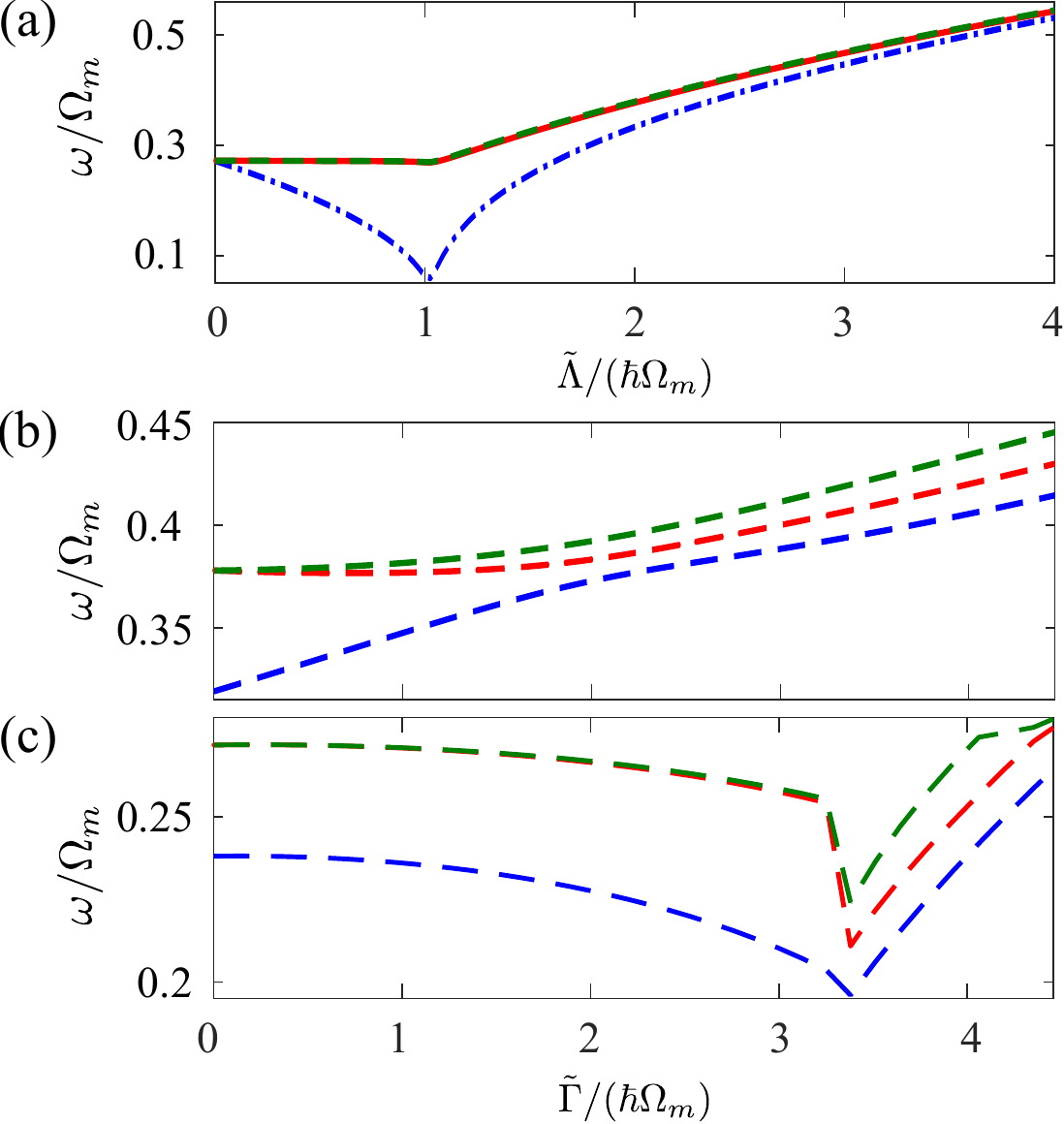}
	\caption{\label{fig:exci}
	The lowest three collective excitations across different phase boundaries indicated in Fig.4 with $\tilde{\Gamma}=0.5\hbar\Omega_m$ (a), $\tilde{\Lambda}=2\hbar\Omega_m$ (b) and $\tilde{\Lambda}=0.25\hbar\Omega_m$ (c) respectively.
	}
\end{figure}

The stability of different phases can be illustrated from their typical excitation spectra.
Fig. \ref{fig:exci} depicts the lowest three collective excitations across the phase boundaries (detailed derivation can be found in Appendix \ref{sec:bog}).
The spectra exhibit non-analytical behaviors when $\tilde{\Gamma}$ or $\tilde{\Lambda}$ sweep across the transition points, as shown in Fig.~\ref{fig:exci}(a) and (c), which indicates the onset of the phase transitions.
By contrast, for $\tilde{\Lambda} > \tilde{\Lambda}_{c}$, the crossover from a periodic LO to a quasi-periodic lattice like DWO is characterized by continuous changes of these excitations, which is also  consistent with the previous discussions, as shown in Fig.~\ref{fig:exci}(b).

\section{\label{sec:con}Experimental consideration and Conclusion}

We note that the considered steady state of the condensates has the potential to be observable within current experimental setup.
As an example, we calculate the relevant parameters for $^{87}$Rb condensate.
For laser beam with the wavelength $\lambda_l=780$nm, the recoil energy can be estimated as
\bea
\hbar\omega_R=\frac{\hbar^2k^2_l}{2m}=\hbar\cdot2\pi\times 3.77~\text{kHz}.
\eea
Here $m$ is the mass of the atom, $k_l =2\pi/\lambda_l$.
The coupling $\lambda_m$  depends closely on the laser power and the cavity finesse, and $\lambda_a$ is determined by the atom-laser coupling strength and detuning (see Appendix \ref{sec:exppara} for details). For typical parameters used in \cite{vochezer2018light}, the relative strength between $\Gamma$ and $\Lambda$ is estimated as $\Gamma/\Lambda=\lambda_a/\lambda_m\approx 3.4\times 10^{-3}$. At first glance, it seems that we can safely ignore the effects of $\hat{H}_{lr}$ for short-time dynamics. However, in the steady state case, the system is determined by the effective
membrane-atom coupling $\tilde{\Lambda}$ and atom-atom interaction $\tilde{\Gamma}$ with
\bea
\tilde{\Lambda} = \frac{\beta\lambda^2_m\lambda^2_a}{4\hbar^2\hbar\Omega_m}N, \mbox{ and }
	\tilde{\Gamma} =\frac{\lambda^2_a}{2\hbar}N.
\eea
The relative strength between $\tilde{\Lambda}$ and $\tilde{\Gamma}$ is then determined by $\lambda_m$, $\beta$, and $\Omega_m$ respectively.
If we set $\beta=2\Omega^2_m/(\Omega^2_m+\gamma^2)=200/101$ and the total particle number $N=10^6$.
The above parameters can then be estimated as
\begin{align}
	\tilde{\Lambda}\approx0.06\hbar\omega_R,~~
	\tilde{\Gamma}\approx4.95\hbar\omega_R,~~
	\tilde{\Gamma}/\tilde{\Lambda}\approx87.87.
\end{align}
Therefore, the induced effective interaction $\hat{\Gamma}$ can be much larger than $\hat{\Lambda}$.
In addition, the lattice potential $V_L$ outside the cavity can also be tuned almost independently.
This can be achieved, for example, by introducing another laser  which is slightly misaligned with the former one.
The two lasers share the same frequency but their relative strengths and phases can be tuned at will.
Therefore, $V_L$ can be changed in a wide range of parameters, as required.
This indicates that the predict phase transition should be attainable within current setup.


To summarize, we have derived explicitly the cavity-mediated non-uniform global atom-atom interaction potential, and studied its effect in a hybrid atom-optomechanical system.
In the steady-state approximation and deep lattice limit, the presence of such global interaction breaks the intrinsic Z$_2$ symmetry induced by membrane-atom coupling, where a right-moved lattice states is favored.
In addition, the non-local properties of such atom-atom interaction can also lead to the breakdown of lattice order, where a self-organized lattice-like state with modified on-site occupations is featured.
The stabilities of these phases are also investigated by solving their Bogoliubov excitations.
The predicted phases provide new possibilities of exploring novel symmetry-breaking physics in this hybrid atom-optomechanical system, and also open up new avenues of research for various exotic quantum states induced by the long-range atom-atom interactions.

\begin{acknowledgments}
	This work was funded by National Natural Science Foundation of China (Grants No. 11474266, No. 11574294, and No. 11774332), the major research plan of the NSFC (Grant No. 91536219), the National Plan	on Key Basic Research and Development (Grant No. 2016YFA0301700), and the ``Strategic Priority Research Program (B)" of the Chinese Academy of Sciences (Grant No. XDB01030200). HP acknowledges support from the US NSF and the Welch Foundation (Grant No. C-1669).
\end{acknowledgments}

\appendix

\section{\label{sec:ham} derivation of the effective Hamiltonian carrying non-uniform global interaction }
In this section, we derive the effective Hamiltonian from original Hamiltonian in the main-text.
Following \cite{vogell2013cavity,vogell2015long,mann2018nonequilibrium,mann2018enhancing,mann2019tuning}, the total Hamiltonian can be written as
\bea
\label{eq:tot}
\hat{H}_{tot} = \hat{H}_m + \hat{H}_a + \hat{H}_l + \hat{H}_{al} + \hat{H}_{ml},
\eea
where $\hat{H}_m$, $\hat{H}_a$, and $\hat{H}_l$ represent the Hamiltonian of membrane, atomic condensate and lasers respectively. $\hat{H}_{al}$ is interaction of atoms with laser beams, and $\hat{H}_{ml}$ describes coupling of mechanical modes with laser fields. The explicit form of these interactions are listed as follows
\bea
\hat{H}_\text{m} &=& \hbar\Omega_\text{m}\hat{a}^{\dagger}\hat{a}, \nn\\
\hat{H}_a &=& \int dz~\hat{\psi}^{\dagger}(z)\mathcal{H}_0\hat{\psi}(z)+\frac{g}{2}\int dz~\hat{\psi}^{\dagger}\hat{\psi}^{\dagger}\hat{\psi}\hat{\psi},  \nn\\
\hat{H}_l&=&\int^{\omega_l+\theta}_{\omega_l-\theta}d\omega~\hbar(\omega-\omega_l)\hat{b}^{\dagger}_\omega \hat{b}_\omega, \nn \\
\hat{H}_{al} &=& \lambda_a\int \frac{d\omega}{\sqrt{2\pi}}(\hat{b}_\omega+\hat{b}^{\dagger}_\omega)\int \hat{\psi}^{\dagger}\sin(z)\sin(\frac{\omega}{\omega_l}z)\hat{\psi} dz,    \nn \\
\hat{H}_{ml} &=& \lambda_m(\hat{a}+\hat{a}^{\dagger})\int \frac{d\omega}{\sqrt{2\pi}}(\hat{b}_\omega+\hat{b}^{\dagger}_{\omega}).  \nn
\eea

\begin{widetext}

According to Heisenberg equation, evolutions of operators $a$, $\psi(z)$ and $b_{\omega}$ read
\bea	
	i\hbar\dot{\hat{a}}&=&[\hat{a},\hat{H}_{\text{tot}}]=\hbar\Omega_m \hat{a}+\lambda_m\int^{\omega_l+\theta}_{\omega_l-\theta}\frac{d\omega}{\sqrt{2\pi}}(\hat{b}_\omega+\hat{b}^{\dagger}_{\omega}),  \label{eq:a} \\	
	i\hbar\dot{\hat{\psi}}(z)&=&[\hat{\psi}(z),\hat{H}_{\text{tot}}]=[\mathcal{H}_0+g\hat{\psi}^{\dagger}(z)\hat{\psi}(z)]\hat{\psi}(z)+ 
	\lambda_a\int^{\omega_l+\theta}_{\omega_l-\theta}\frac{d\omega}{\sqrt{2\pi}}(\hat{b}_\omega+\hat{b}^{\dagger}_{\omega})\sin(z)\sin(\frac{\omega}{\omega_l}z)\hat{\psi}(z),  \label{eq:psi} \\
	i\hbar\dot{\hat{b}}_\omega&=&[\hat{b}_\omega,\hat{H}_{\text{tot}}]=\hbar\Delta_\omega \hat{b}_\omega+\frac{\lambda_m}{\sqrt{2\pi}}(\hat{a}+\hat{a}^{\dagger})+ 
	\frac{\lambda_a}{\sqrt{2\pi}}\int dz \hat{\psi}^{\dagger}(z)\sin(z)\sin(\frac{\omega}{\omega_l}z)\hat{\psi}(z)
\eea
with $\Delta_\omega=\omega-\omega_l$, spectra width $\theta$ of input pulse. The formal solution of $\hat{b}_\omega(t)$ can be written as
\begin{align}
	\label{eq:b}
	\hat{b}_\omega(t)=\hat{b}_\omega(0)e^{-i\Delta_\omega t}+\int^t_0 d\tau e^{-i\Delta_\omega (t-\tau)}\frac{-i}{\hbar\sqrt{2\pi}}
	\left \{ \lambda_m\left( \hat{a}+\hat{a}^{\dagger}\right)_{\tau}+
	\lambda_a\left[ \int dz \hat{\psi}^{\dagger}(z)\sin(z)\sin(\frac{\omega}{\omega_l}z)\hat{\psi}(z)\right]_\tau \right \},
\end{align}
where subscription $\tau$ indicates that the relevant operators is time-dependent.
The first term in Eq.~(\ref{eq:b}) depends on initial condition and can be regarded as a noise.
We substitute Eq.~(\ref{eq:b}) into Eq.~(\ref{eq:a}) and obtain that
\begin{align}	i\hbar\dot{\hat{a}}=\hbar\Omega_m \hat{a}&+\lambda_m\int^{\omega_l+\theta}_{\omega_l-\theta}\frac{d\omega}{\sqrt{2\pi}}\left\{\left[\hat{b}_\omega(0)e^{-i\Delta_\omega t}+\hat{b}^{\dagger}_\omega(0)e^{i\Delta_\omega t}\right]+\int^t_0 d\tau\frac{i}{\hbar\sqrt{2\pi}}\lambda_m(\hat{a}+\hat{a}^\dagger)_\tau\left[e^{i\Delta_\omega(t-\tau)}-e^{-i\Delta_\omega(t-\tau)}\right]\right\} \notag \\
	&+\frac{i}{\hbar}\lambda_m\lambda_a\int^t_0d\tau\int dz~\hat{\psi}^{\dagger}(z)\hat{\psi}(z)\sin(z)\int^{\omega+\theta}_{\omega-\theta}\frac{d\omega}{2\pi}\sin(\omega\frac{z}{\omega_l})\left[e^{i\Delta_\omega(t-\tau)}-e^{-i\Delta_\omega(t-\tau)}\right].
\end{align}
Since we have $\theta\gg\Omega_m$, it is safe to expand the limits of integration $\omega_l\pm\theta$ to $\pm\infty$. The second term relating to $\hat{b}_\omega(0)$ depends on the initial conditions and is known as quantum noises
\begin{align*}
	\hat{F}_a = \int^{+\infty}_{-\infty}\frac{d\omega}{\sqrt{2\pi}}\left[\hat{b}_\omega(0)e^{-i\Delta_\omega t}+\hat{b}^{\dagger}_\omega(0)e^{i\Delta_\omega t}\right] \mbox{ with } \langle \hat{F}_a \rangle=0.
\end{align*}
Using $\int^{+\infty}_{-\infty} e^{i\omega t}d\omega=2\pi\delta(t)$, Heinsenberg equation of the membrane operator $a$ turns into
\begin{align}
	\label{eq:aa}
	i\hbar\dot{\hat{a}}=&\hbar\Omega_m \hat{a}+\frac{1}{2\hbar}\lambda_m\lambda_a\int dz~\hat{\psi}^{\dagger}(z)\hat{\psi}(z)\sin(z)\int^t_0d\tau\int^{+\infty}_{-\infty}\frac{d\omega}{2\pi}\left[e^{i\omega(t-\tau+\frac{z}{\omega_l})}e^{-i\omega_l(t-\tau)}-e^{-i\omega(t-\tau-\frac{z}{\omega_l})}e^{i\omega_l(t-\tau)}+\text{c.c.}
	\right] \notag \\
	=&\hbar\Omega_m \hat{a}+\frac{1}{2\hbar}\lambda_m\lambda_a\int dz~\hat{\psi}^{\dagger}(z)\hat{\psi}(z)\sin(z)\int^t_0d\tau\left[\delta(t-\tau+\frac{z}{\omega_l})e^{-i\omega_l(t-\tau)}-\delta(t-\tau-\frac{z}{\omega_l})e^{i\omega_l(t-\tau)}+\text{c.c.}\right] \notag \\
	=&\hbar\Omega_m \hat{a}-\Lambda\int dz~\hat{\psi}^{\dagger}(z)\sin(2z)\hat{\psi}(z)
\end{align}
with $\Lambda=\lambda_m\lambda_a/(2\hbar)$, where we have neglected the noise term and assumed $z>0$.
Similarly, after substituting Eq.~(\ref{eq:b}) into Eq. (\ref{eq:psi}), we obtain that
\begin{align}
	i\hbar\dot{\hat{\psi}}(z)=[\mathcal{H}_0+g\hat{\psi}^{\dagger}(z)\hat{\psi}(z)]\hat{\psi}(z)+\frac{i}{\hbar}\int^t_0d\tau\int^{+\infty}_{-\infty}\frac{d\omega}{2\pi}(e^{i\Delta_\omega(t-\tau)}-e^{-i\Delta_\omega(t-\tau)})\Big\{\lambda_a\lambda_m(\hat{a}+\hat{a}^{\dagger})_\tau+\notag \\
	\lambda^2_a\big[\int dz^{\prime}\hat{\psi}^{\dagger}(z^{\prime})\sin(z^{\prime})\sin(\frac{\omega}{\omega_l}z^{\prime})\hat{\psi}(z^{\prime})\big]_\tau\Big\}
		\sin(z)\sin(\frac{\omega}{\omega_l}z)\hat{\psi}(z).
\end{align}
Following the same steps in Eq.~(\ref{eq:aa}), we can easily see that for membrane-atom coupling
\begin{align}	i\int^t_0d\tau\int^{+\infty}_{-\infty}\frac{d\omega}{2\pi}\Big[e^{i\Delta_\omega(t-\tau)}-e^{-i\Delta_\omega(t-\tau)}\Big](\hat{a}+\hat{a}^{\dagger})_\tau \sin(\frac{\omega}{\omega_l}z)=\cos(z)(\hat{a}+\hat{a}^{\dagger}),
\end{align}
and for atom-atom coupling
\begin{align}	&\int^t_0d\tau\int^{+\infty}_{-\infty}\frac{d\omega}{2\pi}\Big[e^{i\Delta_\omega(t-\tau)}-e^{-i\Delta_\omega(t-\tau)}\Big]\sin(\frac{\omega}{\omega_l}z)\sin(\frac{\omega}{\omega_l}z^{\prime}) \notag \\	=&-\frac{1}{4}\int^t_0d\tau\int^{+\infty}_{-\infty}\frac{d\omega}{2\pi}\Big[e^{i(\omega-\omega_l)(t-\tau)}-e^{-i(\omega-\omega_l)(t-\tau)}\Big](e^{i\omega\frac{z}{\omega_l}}-e^{i\omega\frac{z}{\omega_l}})(e^{i\omega\frac{z^{\prime}}{\omega_l}}-e^{i\omega\frac{z^{\prime}}{\omega_l}}) \notag \\	=&-\frac{1}{4}\int^t_0d\tau\int^{+\infty}_{-\infty}\frac{d\omega}{2\pi}\Big\{\big[(e^{i\omega(t-\tau+\frac{z}{\omega_l}+\frac{z^{\prime}}{\omega_l})}-e^{i\omega(t-\tau-\frac{z}{\omega_l}+\frac{z^{\prime}}{\omega_l})}+e^{i\omega(t-\tau-\frac{z}{\omega_l}-\frac{z^{\prime}}{\omega_l})}-e^{i\omega(t-\tau+\frac{z}{\omega_l}-\frac{z^{\prime}}{\omega_l})}\big]e^{-i\omega_l(t-\tau)}-\text{c.c.}\Big\} \notag \\	=&-\frac{1}{4}\int^t_0d\tau\big\{\big[\delta(t-\tau-\frac{z^{\prime}+z}{\omega_l})-\delta(t-\tau-\frac{\vert z^{\prime}-z\vert}{\omega_l})\big]e^{-i\omega_l(t-\tau)}-\text{c.c.}\big \} \notag \\
=&\frac{i}{2}\big(\sin(z^{\prime}+z)-\sin\vert z^{\prime}-z\vert\big),
\end{align}
in which we have assumed that $z,z^{\prime}>0$. Combining the above two equations gives that
\begin{align}
	\label{eq:psip}
	i\hbar\dot{\hat{\psi}}(z)=&\Big\{\mathcal{H}_0+g\hat{\psi}^{\dagger}\hat{\psi}-\Lambda(\hat{a}+\hat{a}^{\dagger})
	\sin(2z)-\Gamma\int dz^{\prime}\hat{\psi}^{\dagger}(z^{\prime})\sin(z^{\prime})\hat{\psi}(z^{\prime}) 
	[\sin(z^{\prime}+z)-\sin\vert z^{\prime}-z\vert]\sin(z)\Big\}\hat{\psi}(z)
\end{align}
with $\Gamma=\lambda^2_a/(2\hbar)$.
These two equations Eq.~(\ref{eq:aa}) and Eq.~(\ref{eq:psip}) allow us to write down the effective Hamiltonian $H_{\text{eff}}$ in the main text.

\section{\label{sec:effpotential} effective chemical potential in the self-organized lattice-like phase}
The effective potential at position $z_j$ reads
\bea
	\mathcal{V}(z_j)=\frac{\partial}{\partial n_j}E_{lr}=-\tilde{\Gamma}\Big[\sin^2(z_j)\sum_{k=j}^{L}n_k\sin(2z_k) +\sum_{k=1}^{j-1}n_k\sin^2(z_k)\sin(2z_j)\Big].
\eea
To show the site-dependent feature of $\mathcal{V}(z_j)$, we assume an homogeneous density distribution with $n_j= 1/L$ for all $j=1,\cdots,L$. Therefore, $\mathcal{V}(z_j)$ can be recast into
\bea
\mathcal{V}(z_j) \simeq -\tilde{\Gamma} \frac{1}{2L}\Big[ (1-\cos\xi_j)\sin\xi_j + (1-\cos\xi_j)\sum_{k=j+1}^{L}\sin\xi_k + \sin\xi_j \sum_{k=1}^{j-1}(1-\cos\xi_k)   \Big],
\eea
where we have set
\bea
\xi_k =2 \bar{z}_k = 2[\bar{z}_1+ (k-1)\Delta \bar{z}] = 2[\frac{\pi}{2}+ (k-1)(\pi - \frac{1}{L-1}\frac{\pi}{4})] = \pi + 2(k-1)\pi - \eta_k
\eea
with $\eta_k = (k-1)\pi/2(L-1)$.  Since $\eta_k \in (0,\pi/2)$, when $L\rightarrow \infty$, we can approximate the above summation into integral
\bea
\mathcal{V}(z_j) &\simeq& - \frac{\tilde{\Gamma}}{2L}\Big[(1+\cos\eta_j)\sin\eta_j + (1+\cos\eta_j)\sum_{k=j+1}^{L}\sin\eta_k + \sin\eta_j \sum_{k=1}^{j-1}(1+\cos\eta_k)   \Big] \nn \\
&\simeq& - \frac{\tilde{\Gamma}}{2L}\Big[(1+\cos\eta_j)\sin\eta_j + (1+\cos\eta_j)\frac{1}{\Delta \eta}\int_{\eta_j}^{\pi/2}d\eta\sin\eta + \sin\eta_j \frac{1}{\Delta \eta}\int_0^{\eta_j}d\eta(1+\cos\eta)   \Big]  \nn \\
&=& - \frac{\tilde{\Gamma}}{2L}\Big[ (1+\cos\eta_j)\sin\eta_j + \frac{\cos\eta_j+\eta_j\sin\eta_j+1}{\Delta \eta} \Big]
\eea
with $\Delta \eta = \pi/2(L-1)$. This results in
\bea
\mathcal{V}(z_j)  \stackrel{L\rightarrow \infty}{\longrightarrow} - \frac{\tilde{\Gamma}}{\pi} (\cos\eta_j+\eta_j\sin\eta_j+1).
\eea
Since the function $f(\eta) = \cos\eta+\eta\sin\eta$ increase monotonically as
\bea
\frac{\partial f }{\partial \eta} = -\sin\eta + \sin\eta + \eta\cos\eta = \eta\cos\eta  >0
\eea
when $\eta \in (0,\pi/2)$, we conclude that the effective potential decreases along with the increase of lattice indices $j$.


\section{\label{sec:weaklong}first order phase transitions induced by global non-uniform interaction}
The presence of weak long-range interaction $\tilde{\Gamma}$ not only breaks the intrinsic Z$_{2}$ symmetry but also makes the transition from a lattice order and a right-moved lattice order to be of first order.
To show this, we consider the simplified variational wave-function $
\varphi(z)=\sum_j c_j|z=z^0_j\rangle$
with $z^0_j=j\pi+\delta z$ and $\vert c_j\vert=1/\sqrt{L}$ for all $j \in [1,L]$ ($L$ is the total number of wave packets), then the corresponding energy functional is
\begin{align}
	E(\delta z)=E_0+V\sin^2(\delta z)-\frac{\tilde{\Lambda}}{2}\sin^2(2\delta z)-\frac{\tilde{\Gamma}}{2}\sin^2(\delta z)\sin(2\delta z)
\end{align}
with $E_0$ the remaining interaction energy which is not relevant here.  When $\tilde{\Gamma}/\hbar\Omega_m \ll 1$, around phase boundary $\tilde{\Lambda}\rightarrow\tilde{\Lambda}^c$, we have $\delta z\sim0$ and
\begin{align}
	E(\delta z)\sim p\delta z^2-\tilde{\Gamma}\delta z^3+q\delta z^4+O(\delta z^5)
\end{align}
with $p=V-2\tilde{\Lambda}$, $q=(8\tilde{\Lambda}-V)/3$.
When $\tilde{\Gamma}=0$, The above equation describes a continuous phase transitions at $p=0$ when $q>0$.
Otherwise, the local energy minimal $E(\delta z)$ of can be obtained from
\begin{align}
	E^{\prime}(\delta z)=0~ \Rightarrow ~\delta z_0=0,~\delta z_\pm=(3\tilde{\Gamma}\pm\chi)/(8q)~\text{with}~\chi=\sqrt{9\tilde{\Gamma}^2-32 pq}.
\end{align}
The corresponding energies and second-order derivations are
\begin{alignat}{2}
	&E(\delta z_0)=0, \quad\quad && E^{\prime\prime}(\delta z_0)=2p, \\
	&E(\delta z_{+})=\frac{-1}{2048q^3}(3\tilde{\Gamma}+\chi)^2[\tilde{\Gamma}(3\tilde{\Gamma}+ \chi)-16pq], \quad\quad && E^{\prime\prime}(\delta z_+)=\frac{\chi(\chi+3\tilde{\Gamma})}{8q},\\	
	&E(\delta z_{-})=\frac{-1}{2048q^3}(3\tilde{\Gamma}-\chi)^2[\tilde{\Gamma}(3\tilde{\Gamma}- \chi)-16pq],\quad\quad && E^{\prime\prime}(\delta z_-)=\frac{\chi(\chi-3\tilde{\Gamma})}{8q}.
\end{alignat}
In our case, since 	$\tilde{\Lambda} \rightarrow \tilde{\Lambda}^c = V/2$ and $\tilde{\Gamma} \ll 1$, this ensures $q>0$. Therefore, an overall shift occurs only when $p \le 0$.
This gives the following constrains
\bea
&&\chi\ge3\tilde{\Gamma}>0, \quad\quad\quad \delta z_-<0<\delta z_+,  \\
&&E^{\prime\prime}(\delta z_0)\le0, \quad\quad\quad E^{\prime\prime}(\delta z_+)>0, \quad\quad E^{\prime\prime}(\delta z_-)\ge0.
\eea
Therefore we have $E(\delta z_+)<E(\delta z_-)\le E(\delta z_0)$.
The energy minimal point locates at $\delta z=\delta z_+$ and the ground state is a right-moved lattice phase.
At the critical point $p=0$, the order parameter jumps from zero to its minimal value $\delta z_+\vert_{\text{min}}=3\tilde{\Gamma}/4q$, which indicates that the phase transition is of fist-order.

\section{\label{sec:bog}Bogoliubov excitations}
In this section, we explore the stability and the excitations of different states in the phase diagram.
Taking into account the first-order fluctuations, we rewrite the order parameter as
$ \varphi(z)=\varphi_0(z)+\delta\varphi(z)$,
	where $\varphi_0(z)$ is wavefunction of atomic BEC and $\delta\varphi(z)$ is the fluctuation. Substituting $\varphi(z)$ into the GP equation (Eq. (\ref{eq:gp})), the zero-order term gives the mean-field ground state satisfying
\begin{align}
	i\partial_t\varphi_0(z)=\{\mathcal{H}_0+\tilde{g}\vert\varphi_0(z)\vert^2-\tilde{\Lambda}\kappa[\varphi_0]\sin(2z)
	-\tilde{\Gamma}\chi[\varphi_0,z]\sin(z)\}\varphi_0(z);
\end{align}
For the fluctuation $\delta\varphi(z)$, up to the first-order correction, we get the Bogoliubov equation
\begin{align}
	i\partial_t\delta\varphi(z)=&\Big\{\mathcal{H}_0+2\tilde{g}\vert\varphi_0(z)\vert^2-\tilde{\Lambda}\kappa[\varphi_0]\sin(2z)
	-\tilde{\Gamma}\chi[\varphi_0,z]\sin(z)\Big\}\delta\varphi(z)+\tilde{g}\varphi^2_0(z)\delta\varphi^{\ast}(z) \notag \\
	&-\tilde{\Lambda}\sin(2z)\varphi_0(z)\int dz^{\prime}~\sin(2z^{\prime})\Big[\varphi_0^{\ast}(z^{\prime})\delta\varphi(z^{\prime})+\varphi_0(z^{\prime})\delta\varphi^{\ast}(z^{\prime})\Big] \notag \\
	&-\tilde{\Gamma}\sin(z)\varphi_0(z)\int dz^{\prime}~\sin(z^{\prime})\Big[\sin(z+z^{\prime})-\sin\vert z-z^{\prime}\vert\Big]\Big[\varphi_0^{\ast}(z^{\prime})\delta\varphi(z^{\prime})+\varphi_0(z^{\prime})\delta\varphi^{\ast}(z^{\prime})\Big]
\end{align}
To obtain the Bogoliubov excitation, we rewrite the time-dependent wavefunction as
\begin{align}
	\varphi_0(z,t)=\exp(-i\mu t)\varphi_0(z),~
	\hspace{1cm}\delta\varphi(z,t)=e^{-i\mu t}[e^{-i\omega t}u(z)+e^{i\omega t}\nu^\ast(z)],
\end{align}
with chemical potential $\mu$ and excitation energy $\omega>0$.
Here $\mu$ depends only on wavefunction $\varphi_0(z)$ of condensation
\begin{align}
	\mu=\int dz~\varphi^{\ast}_{0}(z) \Big \{ \mathcal{H}_0+\tilde{g}\vert\varphi_0(z)\vert^2-\tilde{\Lambda} \kappa[\varphi_0]\sin(2z) 
	-\tilde{\Gamma}\chi[\varphi_0,z]\sin(z) \Big \}\varphi_0(z).
\end{align}
The excitation energy $\omega$ is determined by following equations of $u(z)$ and $\nu(z)$
\begin{align}
	\label{eq:ex}
	\omega u(z)=&[\mathcal{A}(z)-\mu]u(z)+[\mathcal{B}(z,z^{\prime})+\mathcal{C}(z,z^{\prime})]u(z^\prime)+[\mathcal{B}^{\prime}(z,z^{\prime})+\mathcal{C}^{\prime}(z,z^{\prime})]\nu(z^\prime)+\tilde{g}\psi^2_0(z)\nu(z), \\
	\label{eq:exc}
	\omega \nu(z)=&-[\mathcal{B}^{\prime\ast}(z,z^{\prime})+\mathcal{C}^{\prime\ast}(z,z^{\prime})]u(z^\prime)-\tilde{g}\psi^{\ast2}_0(z)u(z)-[\mathcal{A}^{\ast}(z)-\mu]\nu(z)-[\mathcal{B}^{\ast}(z,z^{\prime})+\mathcal{C}^{\ast}(z,z^{\prime})]\nu(z^\prime)
\end{align}
with operators
\begin{align}	\mathcal{A}(z)=&\mathcal{H}_0+2\tilde{g}\vert\varphi_0(z)\vert^2-\tilde{\Lambda}\kappa[\varphi_0]\sin(2z)
	-\tilde{\Gamma}^2\chi[\varphi_0,z]\sin(z), \\
	\mathcal{B}(z,z^{\prime})=&-\tilde{\Lambda}\sin(2z)\varphi_0(z)\int dz^{\prime}~\sin(2z^{\prime})\varphi_0^{\ast}(z^{\prime}), \\
	\mathcal{B}^{\prime}(z,z^{\prime})=&-\tilde{\Lambda}\sin(2z)\varphi_0(z)\int dz^{\prime}~\sin(2z^{\prime})\varphi_0(z^{\prime}), \\
	\mathcal{C}(z,z^{\prime})=&-\tilde{\Gamma}\sin(z)\varphi_0(z)\int dz^{\prime}~\sin(z^{\prime})(\sin(z+z^{\prime})-\sin\vert z-z^{\prime}\vert)\varphi_0^{\ast}(z^{\prime}),\\
	\mathcal{C}^{\prime}(z,z^{\prime})=&-\tilde{\Gamma}\sin(z)\varphi_0(z)\int dz^{\prime}~\sin(z^{\prime})(\sin(z+z^{\prime})-\sin\vert z-z^{\prime}\vert)\varphi_0(z^{\prime}).
\end{align}

The presence of membrane-atom and atom-atom couplings brings about nonlocal long-range coupling of excitation modes between $u(z)$ and $\nu(z)$, which is explicitly shown by $\mathcal{B}$, $\mathcal{B}'$ and $\mathcal{C}$, $\mathcal{C}'$ respectively.
From Eq.~(\ref{eq:ex}) and Eq.~(\ref{eq:exc}), we can obtain the excitation spectra using numerical diagonalization. The lowest three excitations are shown in Fig.~\ref{fig:exci} in the main text.
The vanishing imaginary part of the excitations indicates the dynamical stability of all three orders in phase diagram.
The transition between different phases can also be observed from the excitation spectra by their typical analytical behavior around the critical points.


\section{\label{sec:exppara} parameters estimation in a hybrid atom-optomechanical system}
In the main text, we have introduced the dimensionless coordinate $z$ in the total Hamiltonian (see Eq.~(\ref{eq:tot})).
The units of $b_\omega$, $V(or g)$, and $\lambda_{m,a}$  are Hz$^{-1/2}$,   J, and  J$\cdot$(Hz)$^{-1/2}$ respectively.
Following the discussions in \cite{vogell2013cavity}, we can write down the relevant parameters as
\bea
	V&=&\frac{\mu^2\epsilon_{w_l}^2\zeta^2}{\hbar\tilde{\delta}}, \\
	\lambda_a&=&\frac{\sqrt{2\pi}\mu^2\epsilon_{w_l}^2\zeta}{\hbar\tilde{\delta}},\\
	\lambda_m&=&\hbar\frac{\zeta k_l l_a}{\sqrt{\pi}}\vert\tau_m\vert\frac{2\mathcal{F}}{\pi},
\eea
where $\mu$ is the atomic dipole moment, $\epsilon_{w_l}=\sqrt{\frac{\hbar w_l}{\pi\epsilon_0 c S}}$ with the light speed $c$ and the cross-sectional area $S$ of the laser mode, $\zeta$ is related with the laser power $P=\frac{\hbar w_l\zeta^2}{2\pi}$, $\lambda_l$ is the wave-length of laser and $k_l$ wave number of laser, $\tilde{\delta}=w_l-w_{eg}$ is the detuning between the laser frequency $\omega_l$ and the atomic energy gap $\omega_{eg}$,
$l_a=\sqrt{\frac{\hbar}{M\Omega_m}}$ is the characteristic length of the membrane  with mass $M$ and frequency $\Omega_m$, $\tau_m$ is the reflection index,  and $\mathcal{F}$ is the finesse of the cavity.
We assume that a pencil-like shape condensate resides in a potential trap which is a harmonic trap in $x, y$ directions with high frequency $w_x$, $w_y$ and a square well in $z$ axis with length $L_z$.
Other needed physical constants are
\begin{alignat*}{2}
  & \hbar=6.626\times 10^{-34}/(2\pi) ~\text{J}\cdot\text{s}, \quad\quad &&
	\epsilon_0=8.854\times10^{-12}  ~\text{C}/(\text{N}\cdot\text{m}^2), \nn \\
  & c=3\times 10^8 ~ \text{m}/\text{s}, \quad\quad&&
	\mu=3.584\times 10^{-29} ~\text{C}\cdot\text{m}. \nn
\end{alignat*}
In this quasi-one dimensional system, wavefunction of atoms can be assumed as $\psi(\mathbf{r}^\prime)=\psi_g(x^\prime)\psi_g(y^\prime)\psi^\prime(z^\prime)$ with
\begin{align}
	\psi_g(\gamma)=\frac{1}{\sqrt{a_\gamma\sqrt{\pi}}}\exp(-\frac{\gamma^2}{2a_\gamma^2}), ~
	a_\gamma=\sqrt{\frac{\hbar}{m w_\gamma}}, \mbox{ and }
	\gamma=x^\prime,y^\prime.
\end{align}
Effective s-wave interaction can be derived as
\begin{align}
	\frac{g^{\prime}}{2}\int d\mathbf{r}^\prime~\vert\psi(\mathbf{r}^\prime)\vert^4=\frac{g^{\prime}}{2}\int dx^\prime~\vert\psi_g(x^\prime)\vert^4\int dy^\prime~\vert\psi_g(y^\prime)\vert^4\int dz^{\prime}~\vert\psi^\prime(z^{\prime})\vert^4
	\equiv\frac{g}{2}\int dz~\vert\psi(z)\vert^4 \mbox{ with } g=\frac{2\hbar^2a_s}{ma_xa_y\lambda_l}.
\end{align}
where we have used $g^{\prime}=4\pi\hbar^2a_s/m$, $z^{\prime}=z\lambda_l$ and $\psi(z)=\sqrt{\lambda_l}\psi^\prime(z^\prime)$.

We next calculate these parameters by taking $^{87}$Rb atom as an example. The mass of  a $^{87}$Rb atom is $m=87\times1.66\times10^{-27}$ Kg. In experiment \cite{vochezer2018light}, the relevant parameters of the membrane and cavity are
\begin{align}
	M=117 \text{ng}, ~ \Omega_m=2\pi\times 276 \text{kHz}, ~ \tau_m=0.41, ~\mathcal{F}=570.
\end{align}
Here the wavelength of laser is $\lambda_l=780$nm, $\tilde{\delta}=-2\pi\times 1$GHz is the detuning, and the laser power is $P=3.4$mW. The beam waist of laser reads $w_r=250\mu$m, from which we have that $S=\pi w_r^2$.
The frequencies of the harmonic traps can be set as $\{w_x, w_y\}=2\pi\times\{62,85\}$ Hz. Their corresponding characteristic length are $a_x=1.37 \mu\text{m}$, $a_y=1.17 \mu\text{m}$.
We also set the length of the quasi-one dimensional condensate as $L_z=10\lambda_l=7.8 \mu$m $\gg \lambda_l/2$. Using these setting, we can then calculate the recoil energy as
\bea
\hbar\omega_R=\frac{\hbar^2k^2_l}{2m}=2.498\times10^{-30}\text{J}=\hbar\cdot2\pi\times 3.77~\text{kHz}
\eea
 and
\bea
	\lambda_m &\approx & 0.00595941~\text{s}^{1/2}\cdot \hbar\omega_R, \nn \\
	\lambda_a & \approx &0.00002044~\text{s}^{1/2}\cdot \hbar\omega_R, \nn \\
	\frac{\lambda_a}{\lambda_m} & \approx & 3.4\times10^{-3}.\nn
\eea

Next we calculate effective membrane-atom coupling $\tilde{\Lambda}$ and long-range atom-atom interaction $\tilde{\Gamma}$ with
\bea
\tilde{\Lambda} &=& \frac{\beta\Lambda^2}{\hbar\Omega_m}N=\frac{\beta\lambda^2_m\lambda^2_a}{4\hbar^2\hbar\Omega_m}N, \nn \\
	\tilde{\Gamma} &=& \frac{\Gamma}{2} N=\frac{\lambda^2_a}{2\hbar}N.
\eea
The relative strength between $\tilde{\Lambda}$ and $\tilde{\Gamma}$ is then determined by $\lambda_m$, $\beta$, and $\Omega_m$ respectively.
If we set $\beta=2\Omega^2_m/(\Omega^2_m+\gamma^2)=200/101$ and the total particle number $N=10^6$.
Using the above parameters, we can obtained that
\begin{align}
	\tilde{\Lambda}\approx0.06\hbar\omega_R,~~
	\tilde{\Gamma}\approx4.95\hbar\omega_R,~~
	\tilde{\Gamma}/\tilde{\Lambda}\approx87.87.
\end{align}
This ratio indicates that the predict phase transition should be attainable within current setup.

\end{widetext}


\bibliography{v8}

\end{document}